\documentclass[letterpaper,twocolumn,10pt]{article}
\usepackage{usenix-2020-09-draft}

\usepackage{tikz,hyperref}
\usepackage{amsmath,comment,sansmath}
\usepackage[T1]{fontenc}
\usepackage{caption, subcaption, tikz, yfonts, booktabs,multirow,xspace,enumitem}
\usepackage[multiple]{footmisc}
\usepackage{filecontents}
\usepackage[ruled,linesnumbered]{algorithm2e}
\usepackage{xfp} %

\usepackage{listings}
\usepackage{xcolor,breakurl}
\usepackage{dcounter}%
\countstyle{section}%
\DeclareDynamicCounter{algocf}%

\definecolor{codegreen}{rgb}{0,0.6,0}
\definecolor{codegray}{rgb}{0.5,0.5,0.5}
\definecolor{codepurple}{rgb}{0.58,0,0.82}
\definecolor{backcolour}{rgb}{0.95,0.95,0.92}

\lstdefinestyle{mystyle}{
    backgroundcolor=\color{backcolour},   
    commentstyle=\color{codegreen},
    keywordstyle=\color{magenta},
    numberstyle=\tiny\color{codegray},
    stringstyle=\color{codepurple},
    basicstyle=\ttfamily\footnotesize,
    breakatwhitespace=false,         
    breaklines=true,                 
    captionpos=b,                    
    keepspaces=true,                 
    numbers=left,                    
    numbersep=5pt,                  
    showspaces=false,                
    showstringspaces=false,
    showtabs=false,                  
    tabsize=2
}

\newcommand{\Cocoa}{Sync+Sync\xspace}

\newcommand*\circled[1]{\tikz[baseline=(char.base)]{
		\node[shape=circle,draw,inner sep=1pt] (char) {#1};}}

\begin{document}

\date{}

\title{\Large \bf  \Cocoa: A Covert Channel Built on {\tt fsync} with Storage}

\author{
{\rm Qisheng Jiang ~~~~~~~~~~~~~~~~ Chundong Wang}\thanks{C. Wang is the corresponding	author (cd\_wang@outlook.com).}\\
ShanghaiTech University, Shanghai, China
} %

\maketitle

\begin{abstract}
	Scientists have built a variety of covert channels %
	for secretive information transmission %
	with
	CPU cache and main memory. %
	In this paper, %
	we turn to a lower level in the memory hierarchy, i.e., persistent storage.
	Most programs store intermediate or eventual results in the form of files %
	and	some of them call {\tt fsync} to synchronously persist a file with storage device for 
	orderly persistence.
	Our quantitative study  
	shows that one program would undergo significantly longer response time for  {\tt fsync} call 
	if the other program is concurrently calling {\tt fsync}, 
	although they do not share any data. %
	We further find that, concurrent {\tt fsync} calls contend at multiple levels of storage stack %
	due to sharing software structures (e.g., Ext4's journal) and hardware resources (e.g., disk's I/O dispatch queue).
	
	We accordingly build a covert channel named \Cocoa.
	\Cocoa delivers %
	a transmission bandwidth of  
	 20,000 bits per second at an error rate of
	about 0.40\% with an ordinary solid-state   drive.
	\Cocoa can be conducted %
	in cross-disk partition, cross-file system, cross-container, cross-virtual machine, and even cross-disk drive
	fashions,
	 without sharing %
	data  between programs.
	Next, we launch side-channel attacks with \Cocoa and manage to 
	precisely detect operations %
	of a victim database (e.g., insert/update %
	 and B-Tree node split). 
	We also leverage \Cocoa to distinguish 
	applications and websites with high accuracy by %
	 detecting and analyzing
	their {\tt fsync} frequencies and flushed data volumes.
	These attacks are useful
	 to support further fine-grained information leakage. %
\end{abstract}

\section{Introduction}\label{sec:intro}

Computer scientists have explored a variety of covert channels to leak information.
CPU cache and memory are main building blocks for many existing covert channels,
as programs share them 
when concurrently running. 
In this paper, 
we study a lower layer in the memory hierarchy, i.e.,
persistent storage and file system, to discover a new covert channel.

The file is the most general form in which programs persistently store their intermediate or eventual execution results. 
There are different I/O models for programs to consider, including %
buffered I/O, direct I/O, and synchronous I/O with {\tt fsync}~\cite{FS:rethink-sync:OSDI-2006,FS:WALDIO:ATC-2015,FS:BarrierFS:FAST-2018,FS:fsync-failure:ATC-2020,FS:exF2FS:FAST-2022}.
We focus on %
synchronous I/O with {\tt fsync}
for two reasons. Firstly,
many applications employ 
{\tt fsync} %
for orderly durability and consistency, such as databases and mail 
services~\cite{journal:iJournaling:ATC-2017,FS:fsync-failure:ATC-2020}.
Secondly, compared to buffered I/O and direct I/O that write data to 
the page cache of operating system (OS) and internal device cache of storage device,
 respectively,
 {\tt fsync} has a more deterministic and constant %
 timing.
In short,
when a program calls {\tt fsync} on a particular 
file, %
 file system (e.g., Ext4) 
flushes dirty data buffered in memory 
pages as well as file metadata (e.g., {\tt inode}),
for the file through the block I/O ({bio}) layer
 to underlying block device, such as a hard disk drive (HDD) 
 or solid-state drive (SSD). 
To forcefully persist the file, {\tt fsync}
carries bio flags (i.e., {\tt REQ\_PREFLUSH}
and {\tt REQ\_FUA}) to instruct storage device to flush the internal cache.
Different file systems   have different implementations for {\tt fsync}.
Take the widely-used Ext4 for example. %
Ext4 is a journaling file system generally mounted in the default {\tt data=ordered} mode.
Upon an {\tt fsync},
it persistently commits file metadata to an on-disk journal
after persisting file data~\cite{journal:iJournaling:ATC-2017}.
Also, a storage device has mechanisms to guarantee orderly persistence
for each {\tt fsync}~\cite{FS:BarrierFS:FAST-2018,FS:OPTR:ATC-2019}.

With regard  to the increasingly large capacity of terabytes or even more
per disk drive, multiple programs are likely to run and share the same disk in a local machine
or cloud dedicated server~\cite{security:cgroups-sync:CCS-2019}.
We find that, whenever two programs stay 
in the same disk partition with the same file system or different disk partitions
with different file systems, one of them ({\em receiver}/{\em attacker}) would 
undergo much longer response time to wait for the return of  {\tt fsync} 
if the other program ({\em sender}/{\em victim}) is concurrently 
calling {\tt fsync}. For example,
with Ext4 mounted on an ordinary SSD, program   X  has to wait more than twice the time
($\frac{\textnormal{43.13}\mu s}{\textnormal{21.39}\mu s}$)
for the completion of {\tt fsync}
when the other program Y is simultaneously calling {\tt fsync}. 
Note that X and Y
operate with absolutely unrelated files, without any on-disk file or in-memory data 
being shared.

The significant impact of {\tt fsync} is due to contention at 
multiple levels in the storage stack.
When two programs are running within the same file system, 
they share structures of both file system and storage device.
For example, the aforementioned  journal of Ext4 
is a globally shared structure Ext4 uses to
 record changes of files 
in the unit of transactions. 
The  journal is an on-disk circular buffer and transactions must be sequentially committed to it~\cite{journal:iJournaling:ATC-2017,FS:journaling-cores:FAST-2018,journal:Z-journal:ATC-2021}.
By default, Ext4 maintains one running transaction to take in modified metadata blocks for files.
An {\tt fsync} targeting a file explicitly 
 commits to the journal
 the current running transaction where the file's {\tt inode} block is placed. 
The {\tt fsync} may need to wait for the completion of committing previous  transaction and also
  hinder the progress of subsequent transactions.
   As a result, %
   Ext4 forcefully serializes %
   {\tt fsync}s and each {\tt fsync} must stall until Ext4 commits
  the transaction for current one. %
 Although researchers have proposed the fast commit to optimize {\tt fsync}~\cite{journal:iJournaling:ATC-2017,FS:fast-commit-Ext4},
 our study shows that the introduction of it does not alleviate
 the substantial interference between %
 {\tt fsync}s. 
Worse, concurrent
  {\tt fsync}s %
  also share and compete on other
limited software and hardware resources for storage.
The bio and disk 
drivers need to serialize
and write  down metadata and data for each {\tt fsync} through queues. 
Consequently,
 programs running %
 in different disk partitions mounted with different
file systems %
severely suffer from each other's {\tt fsync}s.

These observations motivate us to propose a new covert channel named 
\Cocoa for secretive communication.
With \Cocoa, the sender transmits a bit by calling {\tt fsync} or not
while the receiver receives the bit by calling {\tt fsync}.
For instance, the receiver gets `1' after undergoing a significantly longer response time and receives `0' otherwise. 
As mentioned,
cross-file system and cross-partition \Cocoa channels
are effectual. We also find that receiver and sender staying in
different containers and virtual machines (VM)
can efficiently
communicate with \Cocoa in cross-container and -VM fashions. 
To establish a reliable and efficient covert channel,
we only need both sides 
co-located
in the same storage device.
In addition,
\Cocoa does not demand sender and receiver to share on-disk or in-memory data. Both 
just  call {\tt fsync}s on absolutely
unrelated files in the user space, as {\tt fsync} is an unprivileged system call. 
Our further exploration  shows that cross-disk \Cocoa is also functional.
This enables \Cocoa to gain high flexibility and viability.

 In practical, %
databases widely employ {\tt fsync}s for %
durability and consistency.
Many applications also use {\tt fsync}s to store data.
We hence use \Cocoa to figure out sensitive information from a victim %
that calls {\tt fsync} over time. %
For example, we leverage \Cocoa to identify the runtime operations, such as insert, update, and 
B-Tree node split, for SQLite~\cite{app:SQLite}. %
We also differentiate  access patterns between common applications 
and websites 
with \Cocoa.
One example is that  an attacker 
can leverage \Cocoa side channel to distinguish 
Facebook and Twitter with 
100\%
accuracy by analyzing their I/O traces.
\Cocoa hence provides an effective tool to observe a victim's I/O %
characteristics.
Finally, we perform a keystroke attack with \Cocoa to explore the sensitivity to user inputs with an accuracy of about 99.2\%.

Because of the necessity and prevalence of {\tt fsync}s,
 attackers calling {\tt fsync}s with \Cocoa are difficult %
 to be discovered.
The defense against \Cocoa is also not straightforward. The performance
cost and interference of {\tt fsync} is a classic issue and it is impossible
to avoid {\tt fsync}s regarding critical persistence needs of applications.
Also, existing techniques to reduce interference for {\tt fsync} such as the aforementioned fast commit 
 for Ext4 are still vulnerable to \Cocoa. %
Though, we have given few  suggestions %
to mitigate the impact of
\Cocoa attacks.

To sum up, we make the following contributions.
\begin{itemize}
	\item We reveal and build a timing-based  covert channel named \Cocoa
	 at the persistent storage  with {\tt fsync} system call.
	To our best knowledge, \Cocoa is the first covert channel 
	that makes use of {\tt fsync} %
	at the persistent storage   without sharing any data between sender  and receiver.
	\item We quantitatively %
	demonstrate that the %
	 covert channel  of \Cocoa
	achieves a transmission
	 bandwidth of 20,000 bits per second (bps) with about 0.40\% error rate.
 \Cocoa covert channel effectively works in cross-partition, cross-file system, cross-container,  cross-VM, and cross-disk fashions. 
We also introduce various noise to test and justify the robustness of \Cocoa.
	\item We perform side-channel attacks to target real-world application programs that use {\tt fsync}s in their implementations. %
	 \Cocoa is able to classify and identify   
	 database operations %
	such as insert, update, and B-Tree node split with SQLite.
	It  further differentiates  applications and websites from their behaviors of calling {\tt fsync}s with varying frequencies, timings, and data volumes. 
	A keystroke attack is also practicable. %
	\Cocoa accomplishes all these attacks  with high accuracy.
\end{itemize}

The rest of this paper is organized as follows. We 
study  related works and background knowledge for side-channel attacks in Section~\ref{sec:bg}.
We illustrate
why {\tt fsync} makes an effectual covert channel in Section~\ref{sec:bg:fsync}. 
We  present the attack model and
effectiveness of \Cocoa in communicating between sender and receiver in Section~\ref{sec:covert:channel}. 
In Section~\ref{sec:side:channel}, we detail how \Cocoa is used to perform concrete attacks for information leakage. We show 
discussions and defenses for \Cocoa in Section~\ref{sec:discussion}. 
We conclude the paper in Section~\ref{sec:conclusion}.

\section{Background and Related Work}\label{sec:bg}

\subsection{Side-Channel Attacks}\label{sec:related}

Side-channel attacks pose a significant threat to computer systems and software security.
They exploit  information leakage through various channels, including shared hardware resources, system software and hardware, applications, and other observations. 
In this section, we discuss notable side-channel attacks that are both general and contention-related.

{\bf 1) Shared Hardware Resources}: 
Shared hardware resources, such as the CPU cache~\cite{security:Prime+Probe:CT-RSA-2006,security:Prime+Probe:SP-2015,security:Flush+Reload:Security-2014,security:Prime+Scope:CCS-2021,cpu:NetCAT,security:prefetch+reload:SP-2022}, 
translation lookaside buffer (TLB)~\cite{tlb:TLBleed}, 
branch predictor~\cite{branch:BranchScope,branch:JumpOverASLR},
persistent memory~\cite{security:NVLeak:USENIX-2023,security:pmem:USENIX-2023}, 
and GPU~\cite{gpu:GPUCC,gpu:renders},
are common targets for side-channel attacks. 
Researchers have developed numerous attacks that exploit contention %
by leveraging these resources. 
For instance, Prime+Probe~\cite{security:Prime+Probe:CT-RSA-2006,security:Prime+Probe:SP-2015},
Flush+Reload~\cite{security:Flush+Reload:Security-2014}, and Prime+Scope~\cite{security:Prime+Scope:CCS-2021} 
utilize cache sets or specific cache lines shared among running programs. 
Besides CPU cache contention, side channels utilizing other shared resources, 
such as the directory for cache coherency, cache bandwidth, and coherency states, 
have also been employed to leak information~\cite{security:cache-coherence:HPCA-2018,security:CacheBleed:CHS-2016,security:directory:SP-2019}.

{\bf 2) System Software and Hardware}: 
Side-channel attacks can also target system software and hardware. 
Lower-level main memory components, such as the OS's page cache and shared page mapping with and without page faults, 
have been observed to imply side channels~\cite{security:page-fault:SP-2015,security:page-fault:ASIACCS-2016,security:no-page-fault:Security-2017,security:page-fault:CCS-2017,security:pagecache:CCS-2019}.
Page cache~\cite{security:cgroups-sync:CCS-2019}, file systems~\cite{security:DUPEFS:USENIX-2023}, 
page walker~\cite{pagewake:Binoculars}, just-in-time %
 compilation~\cite{JIT:DeJITLeak}, 
and database 
 queries~\cite{security:database:CCS-2016,security:database:CCS-2018,security:database:SP-2018,security:SQLite-query:Security-2021}
have been exploited for side-channel attacks too. 
For example, 
Gao et al.~\cite{security:cgroups-sync:CCS-2019} demonstrated a covert channel across containers 
by utilizing {\tt sync} to write back all dirty pages in the OS's page cache.
Bacs et al.~\cite{security:DUPEFS:USENIX-2023} introduced a timing-based side channel called DupeFS that utilized inline file system deduplication.
Their observation is that if data an attacker writes have been deduplicated due to a previous
write done by the victim, the attacker would observe a shorter response latency.

{\bf 3) Applications}: 
Web browsers and other applications are also susceptible to side-channel attacks. 
Researchers have developed attacks that target web browsers by exploiting various vulnerabilities~\cite{security:browser-storage:ACSAC-2016,security:browser:USENIX-2021,security:browser:Security-2022}. 
For example, Kim et al.~\cite{security:browser-storage:ACSAC-2016}
utilized the interactions between websites and the disk space quota for different websites to infer visited websites, access history, and login status with a particular website.

{\bf 4) Other Observations}: 
Side-channel attacks have also been launched in other dimensions, such as power consumption analysis~\cite{security:power:Security-2021,power:iknowwhatyousee,power:powerchannels}.
For example, 
Chen et al.~\cite{security:power:Security-2021} proposed a side channel based on the idle power management for CPUs.
By observing the power consumption pattern of a victim, attackers can analyze and deduce sensitive information.

\subsection{I/O Stack in OS}\label{sec:bg:io}

\begin{figure}[t]
	\centering
	\includegraphics[width=\columnwidth]{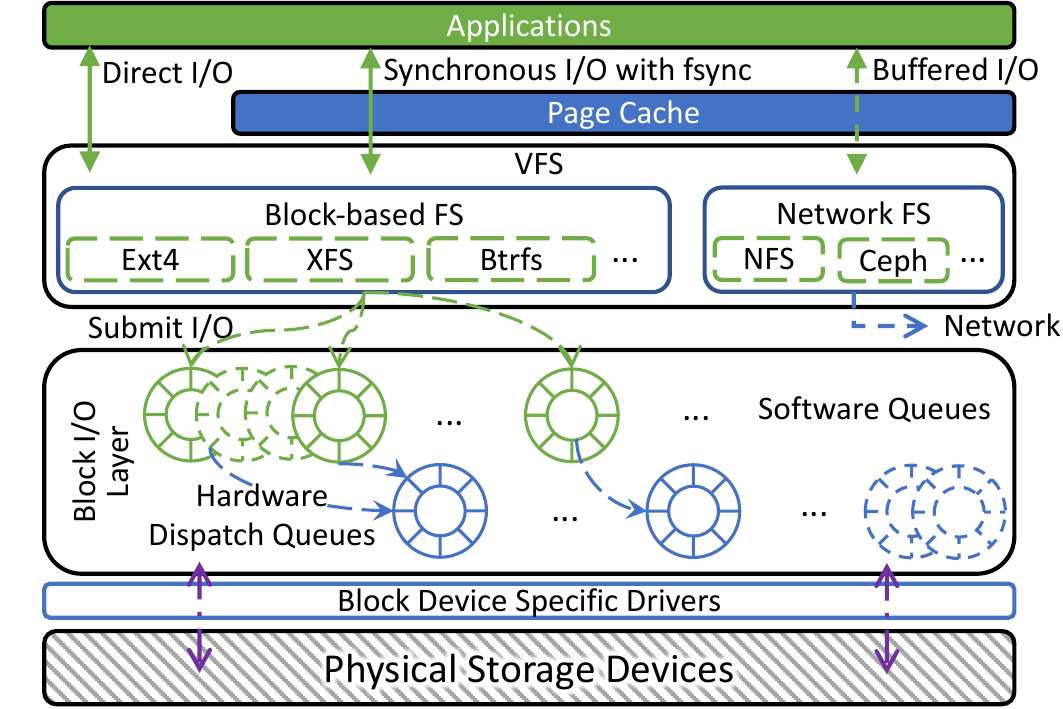}
	\caption{An Overview of Linux I/O Stack.}
	\label{fig:io-stack}
\end{figure}

Most applications store data in the form of files
with file system and 
underlying
persistent storage device, such as SSD or HDD.
Without loss of generality,
we follow Linux I/O stack shown in~\autoref{fig:io-stack} to illustrate how 
OS handles and stores data into block I/O device.
The main layers in Linux's storage stack include
virtual file system (VFS),
 file systems, generic block I/O layer, and drivers for specific storage devices.

\textbf{VFS and File Systems.} %
VFS provides high-level abstractions of file operations for applications to utilize,
such as
{\tt open}, {\tt close}, {\tt write}, {\tt read}, and {\tt fsync}  interfaces.
A particular file system (e.g., Ext4 or XFS) handles  file operations with underlying storage device.
File system is also responsible for organizing and managing storage space 
and guarantees essential properties like consistency and durability for files. 
Occasionally,
 OS may crash due to unexpected events, such as software bugs (e.g., kernel panic)
or power outage. 
File system shall ensure that the modifications onto file metadata and/or data are crash-recoverable 
in line with a consistency level configured on mounting the file system.
Journaling (logging)~\cite{FS:journaling-cores:FAST-2018,journal:Z-journal:ATC-2021,journal:iJournaling:ATC-2017}, 
copy-on-write (CoW)~\cite{FS:Btrfs}, 
and soft updates~\cite{FS:soft-updates} are main techniques that  file systems
adopt to guarantee crash consistency for file metadata (e.g., {\tt inode}) and data.
Let us take Ext4 with journaling for illustration because of the wide deployment of it.
Ext4 maintains an on-disk journal as a ring buffer.
When Ext4 is mounted in the default {\tt data=ordered} mode, only modified   file metadata are 
written to the journal. Ext4 composes a {\em transaction} in memory with multiple metadata blocks
and commits the transaction as a unit to the journal.

Ext4 writes file data and metadata into 
OS's page cache 
to handle a write request via buffered I/O.
It triggers a write-back of these dirty pages
with regard  to several conditions. The first one is a periodical flush that occurs
 every five seconds
by default.
The second one is that 
dirty memory pages take more than %
a proportion of total memory (10\% by default)~\cite{redhat:writeback}.
The third one is an explicit {\tt fsync} received from applications.
In the {\tt data=ordered} mode,
Ext4 strictly writes file data %
to on-disk  blocks allocated to a file
before it commits the file's
metadata block to the journal.
Later, Ext4 checkpoints metadata blocks in place.
The {\tt inode} is the most important metadata for a file and
 contains the file's length, access time, and access permissions.
 Once an {\tt inode} block is committed to the
 journal before a crash happens, Ext4 can recover the file since
 both file data and metadata have been made durable and retrievable.
In addition, Ext4 maintains only one running transaction at runtime.
Hence
all files concurrently share one transaction. 
The aforementioned three conditions transform a running transaction to  
be committing transaction and Ext4 generates the next running transaction.
It orderly commits consecutive transactions 
 to the journal.

\begin{figure*}[htbp]
	\begin{subfigure}{0.24\textwidth}
		\includegraphics[width=\textwidth,page=1]{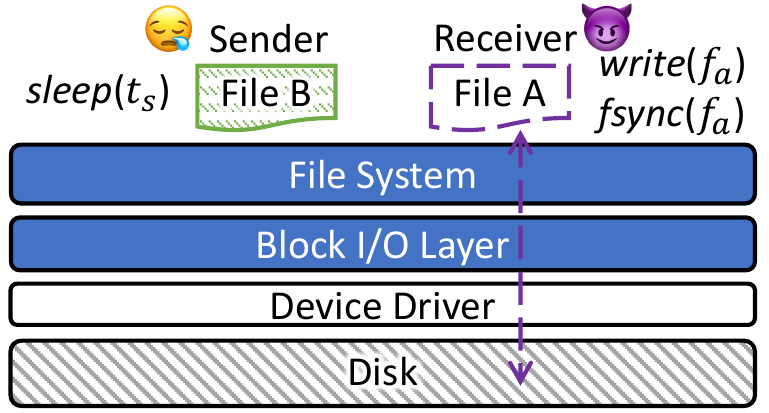}
		\caption{{\tt fsync} without Contention.}
		\label{fig:fsync:wo-contention}
	\end{subfigure}	
	\hfill
	\begin{subfigure}{0.24\textwidth}
		\includegraphics[width=\textwidth,page=2]{transmit.pdf}
		\caption{{\tt fsync} with Contention.}
		\label{fig:fsync:with-contention}
	\end{subfigure}
	\hfill
	\begin{subfigure}{0.24\textwidth}
		\includegraphics[width=\textwidth,page=3]{transmit.pdf}
		\caption{Contention in File System.}
		\label{fig:fsync:journal}
	\end{subfigure}
	\hfill
	\begin{subfigure}{0.24\textwidth}
		\includegraphics[width=\textwidth,page=4]{transmit.pdf}
		\caption{Contention in Block I/O Layer.}
		\label{fig:fsync:bio}
	\end{subfigure}
	\caption{An Illustration of Contention Caused by {\tt fsync}.}
	\label{fig:fsync:contention}
\end{figure*}

\textbf{Block I/O Layer.} %
Block I/O (bio) layer connects a particular file system and underlying storage device.
As shown in the middle of~\autoref{fig:io-stack},
Linux maintains multiple software queues ({\tt blk-mq}) for bio~\cite{bio:multi-queue}.
Each block device contains per-core software staging queues and hardware dispatch queue(s) 
depending on the device's hardware and driver.
File system submits a bio request to 
software staging queues and waits 
for I/O scheduler to dispatch.
Then the device driver
  interacts with device to complete the I/O request.
Although multiple software queues mitigate competitions between I/O requests at the bio layer,
the limited number of hardware queues installed in a block device restricts the device's capability~\cite{FS:BarrierFS:FAST-2018}.
Thus, concurrent 
I/O requests issued to a 
 storage device 
 interfere with each other,
especially for devices with one single hardware dispatch queue, 
such as ordinary HDDs and SSDs.

\section{The Contention Caused by {\tt fsync}}\label{sec:bg:fsync}

OS provides multiple unprivileged system calls,
such as 
{\tt fsync}, {\tt fdatasync}, and {\tt msync},
 for user-space programs %
to synchronously flush data %
to storage device for orderly persistence. %
{\tt fdatasync} and {\tt msync} are variants of {\tt fsync}.
They
either do not flush file metadata or specifically synchronize a memory-mapped file, respectively \cite{FS:rethink-sync:OSDI-2006,FS:msync:Eurosys-2013}.
On receiving an {\tt fsync} call for a file, %
file system transfers all modified in-core data and metadata (i.e., dirty pages in OS's page cache) of the file 
to storage device and issues  an ending 
 bio request %
 with two flags ({\tt REQ\_PREFLUSH} and {\tt REQ\_FUA}) set. 
Then the file's metadata and data are to %
 be forcefully persisted into storage. %
File system returns a success or fail  to the caller program %
according to the device's returned signal.
A successful {\tt fsync} means that
the file's durability is achieved and %
all changed metadata and data become retrievable even if   OS %
suddenly crashes.
Because of the explicit and synchronous persistence of {\tt fsync},
 many applications, particularly ones like databases
that are highly concerned
with data consistency and durability, widely employ {\tt fsync}s in their implementations.

In addition, there is a {\tt sync} system call that flushes {\em all} dirty pages in the OS's page cache
to storage device and applications do not need to specify a file~\cite{security:cgroups-sync:CCS-2019}.
Therefore,
{\tt sync} and {\tt fsync} are like  instructions of flushing all cache lines and one 
particular cache line, i.e., {\tt wbinvd} and {\tt 
	clflush} in the x86 instruction set architecture (ISA), respectively. 
Since {\tt sync} causes significantly longer time than {\tt fsync}
and is much easier to be perceived, we focus on {\tt fsync} %
in this paper.

\subsection{Contention within File System}

As shown in~\autoref{fig:fsync:wo-contention} and~\autoref{fig:fsync:with-contention},
two programs that call {\tt fsync}s interfere each other.
Let us %
 use Ext4 for illustration.
When a program %
 calls {\tt fsync} on a particular file, 
Ext4 flushes data pages and commits the transaction in which the
file stays to the journal. %
As Ext4 maintains only one running transaction
and the journal is organized in a ring buffer,   the {\tt fsync} may need to
wait for the completion of committing previous %
transaction.
Thus, at the Ext4 file system level, 
the shared global journal and serialization of committing transactions 
make one {\tt fsync} undergo much
 longer response latency %
 in the presence of 
 another concurrent {\tt fsync}.

We use the following notations to decompose the {\tt fsync} latency for a certain file A
with and without contention.
$$
\begin{aligned}
    T_{fsync}^{A} &:= \text{total {\tt fsync} latency for the file A.} 
    \\
    T_{data}^{A} &:= \text{latency to flush file A' dirty data pages (blocks).}  \\    
    T_{flush} &:= \text{latency to flush device cache with flush command.} \\
	T_{meta}(\cdot) &:= \text{latency to commit a metadata block.} \\
	\Upsilon_{prev} &:= \text{latency to finish previous committing transaction }\\ 
	& \text{\ \ \ \ \ \ if applicable}.
\end{aligned}
$$
Assuming that the metadata block of file A is the block $\tau_n$ 
in current transaction, 
and blocks $\tau_0$ to $\tau_{n-1}$ are prior elements for $\tau_n$ in the transaction,
we have
\begin{equation}
\tau_0 \succeq \tau_1 \succeq ... \succeq \tau_{n-1} \succeq \tau_n,
\end{equation}
in which $\tau_i\succeq\tau_{i+1}$ means $\tau_i$ {\em happens before} $\tau_{i+1}$ 
(i.e., $\tau_i$ has entered the   transaction earlier than $\tau_{i+1}$).
According to Ext4's {\tt fsync} behavior,
the {\tt fsync} latency of file A,   $T_{fsync}^{A}$, is contributed by the following parts,
\begin{equation}
	T_{fsync}^{A} = T_{data}^{A} + \Sigma_{i = \tau_0}^{\tau_n}{T_{meta}(i)} + T_{flush} + \Upsilon_{prev}.
\end{equation}

Concurrent {\tt fsync}s cause dramatic impact on {\tt fsync} latencies.
For example, 
we write   data to file A without any other program issuing {\tt fsync} and
measure the latency as $T_{fsync1}^{A}$.
Next, 
we do the same {\tt fsync} on file A but, in the meantime, we make 
the other program   perform {\tt fsync}
on a different file B.
We denote the second latency for flushing file A as $T_{fsync2}^{A}$.
As shown in~\autoref{fig:fsync:journal}, two programs orderly proceed with Ext4's journaling
(\circled{1} $\succeq$ \circled{2}).
Our quantitative tests %
 confirm that
$T_{fsync2}^{A}$ is much greater than $T_{fsync1}^{A}$, which is mainly because
$\Upsilon_{prev}$ 
emerges with the interference of syncing file B.

\subsection{Contention within Storage Device}\label{sec:fsync:bio-contention}

 {\tt fsync} demands underlying storage device to synchronously 
write data down and
forcefully drain the device cache for eventual persistence.
Because of the limited hardware resources of a storage device, further contention
 occurs at the device level between concurrent {\tt fsync}s. 
Each {\tt fsync} is transformed to bio requests 
 that orderly flush data and metadata into disk. %
When an I/O request is submitted to the bio layer,
it mainly goes through three 
 stages~\cite{linux:blktrace, bio:multi-queue,FS:BarrierFS:FAST-2018}.
    {\bf 1) Q2I}: The submitted I/O request is preprocessed (e.g., request split and address remapping) 
    and then inserted or merged into a  request 
 queue.
    {\bf 2) I2D}: The I/O request waits in the request queue, staying idle until
    the I/O scheduler dispatches and puts it in the dispatch queue (see \circled{1} in~\autoref{fig:fsync:bio}).
    {\bf 3) D2C}: The I/O request is issued to corresponding device driver for the device to handle and the bio layer 
    waits  at the  {completion} queue for I/O completion (see \circled{2} in~\autoref{fig:fsync:bio}).

In each stage,   I/O requests issued for different files 
compete for resources against each other.
The contention at   the D2C stage
 is even worse  due to the limited capability of storage
device, as overwhelming synchronous I/O requests are likely to engage
 I/O scheduler in %
 spending increasingly longer time in dispatching next ones. %
As shown in~\autoref{fig:fsync:bio},
 there is a high likelihood of contention for {\tt fsync}s at the device level.

Each {\tt fsync} generates
 bio requests with {\tt REQ\_PREFLUSH} and {\tt REQ\_FUA} flags,
 which are eventually translated to device flush commands to 
drain the internal device cache. This
 makes one more %
contention point between {\tt fsync}s at the device level.
For some file system, such as Ext4 or XFS, 
{\tt fsync}s
that do not follow any change of file data or metadata, e.g., with {\tt write}
{\tt ftruncate} operations,
still interfere with each other,
because Ext4 and XFS always issue the device flush command for {\tt fsync}
regardless of a modification received or not onto files.

\begin{table}[t]
    \centering
	\caption{{\tt fsync} Latencies with and without Contention.}\label{tab:fsync-latency}
    \resizebox{\columnwidth}{!}{%
    \begin{tabular}{crrrcrrr}
    \hline
    \begin{tabular}[c]{@{}c@{}}Operation for\\ Measurement \end{tabular} & \begin{tabular}[c]{@{}c@{}}Standalone\\ Latency ($ns$)\end{tabular} & \begin{tabular}[c]{@{}c@{}}Standard\\  Deviation \end{tabular} & \begin{tabular}[c]{@{}c@{}}Standard\\  Error \end{tabular} & \begin{tabular}[c]{@{}c@{}}Operation for\\ Competitor \end{tabular} & \begin{tabular}[c]{@{}c@{}}Contention\\ Latency ($ns$)\end{tabular} & \begin{tabular}[c]{@{}c@{}} Standard\\ Deviation \end{tabular} & \begin{tabular}[c]{@{}c@{}} Standard\\ Error \end{tabular} \\ \hline
    \multirow{3}{*}{{\tt ftruncate}+{\tt fsync}} & \multirow{3}{*}{124725.81} & \multirow{3}{*}{5067.66} & \multirow{3}{*}{506.77} & {\tt ftruncate}+{\tt fsync} & 186635.09 & 4035.34 & 403.53 \\ %
     &  &  &  & {\tt write}+{\tt fsync} & 165046.60 & 20210.83 & 2021.08 \\ %
     &  &  &  & {\tt fsync}-only & 165989.67 & 17878.06 & 1787.81 \\ %
    \multirow{3}{*}{{\tt write}+{\tt fsync}} & \multirow{3}{*}{55707.18} & \multirow{3}{*}{21756.38} & \multirow{3}{*}{2175.64} & {\tt ftruncate}+{\tt fsync} & 104521.34 & 24380.61 & 2438.06 \\ %
     &  &  &  & {\tt write}+{\tt fsync} & 84345.38 & 27756.34 & 2775.63 \\ %
     &  &  &  & {\tt fsync}-only & 102094.32 & 26968.84 & 2696.88 \\ %
    \multirow{3}{*}{{\tt fsync}-only} & \multirow{3}{*}{21390.42} & \multirow{3}{*}{2478.57} & \multirow{3}{*}{247.86} & {\tt ftruncate}+{\tt fsync} & 47428.24 & 21527.58 & 2152.76 \\ %
     &  &  &  & {\tt write}+{\tt fsync} & 51112.34 & 9088.28 & 908.83 \\ %
     &  &  &  & {\tt fsync}-only & 43133.73 & 2521.81 & 252.18 \\ \hline
    \end{tabular}%
    }
\end{table}

\subsection{The Viability of {\tt fsync} Channel}\label{sec:viability}

In order to empirically understand {\tt fsync}, %
we test it with and without contention.
We   make a program X and 
 measure the {\tt fsync} latency without contention under following    
settings 
when program X synchronizes a file
with Ext4 mounted on an ordinary SSD 
(SAMSUNG PM883 SATA SSD). 
\begin{itemize}
\item The {\tt ftruncate} + {\tt fsync} stands for 
synchronizing a file after truncating the file to a random size, which   triggers the commit of file metadata
into Ext4's journal.
\item The {\tt write} + {\tt fsync} represents that 
we call {\tt fsync} on a file after overwriting the file for 1KB with  data contents. 
\item The {\tt fsync}-only means that 
we measure the latency of {\tt fsync} without any modification of file data or metadata. 
\end{itemize}
We only record the latency of {\tt fsync} while the time of writing data
or truncating a file is not counted.
Next, we keep the other program Y running with %
the three settings %
on a different file co-located in the same disk for contention.
We measure latencies of program X under different configurations %
and repeat
each configuration for 100 times.

\begin{figure}[t]
	\begin{subfigure}{0.235\textwidth}
		\includegraphics[width=\textwidth,page=1]{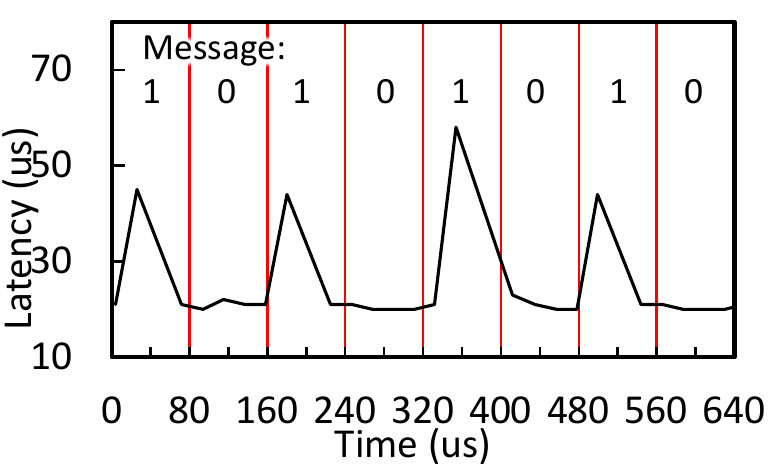}
		\caption{{\tt fsync}-only.}
		\label{fig:raw:fsync-only}
	\end{subfigure}	
	\hfill
	\begin{subfigure}{0.235\textwidth}
		\includegraphics[width=\textwidth,page=2]{raw-ftruncate.pdf}
		\caption{{\tt ftruncate}+{\tt fsync}.}
		\label{fig:raw:ftrucate-fsync}
	\end{subfigure}
	\caption{Raw Traces of Cross-file \Cocoa Channels within Ext4.}
	\label{fig:raw-trace-ftruncate}
\end{figure}

\autoref{tab:fsync-latency} captures the average latency 
for each configuration
after 100 executions.
It is evident that the latency of {\tt fsync} that program X observes
increases significantly when program Y is simultaneously calling {\tt fsync}. 
For example, the latency of {\tt fsync}-only jumps by 
\fpeval{round(47428.24 / 21390.42, 1)}$\times$, 
\fpeval{round(51112.34 / 21390.42, 1)}$\times$, 
and \fpeval{round(43133.73 / 21390.42, 1)}.0$\times$ 
when program Y is concurrently doing
{\tt ftruncate} + {\tt fsync}, {\tt write} + {\tt fsync}, and {\tt fsync}-only, respectively.
As also shown by~\autoref{tab:fsync-latency},
in spite of dramatic increase for latencies,
the standard deviation and standard error
 generally fluctuate in an acceptable and observable range.
Among three settings, 
{\tt write} + {\tt fsync} needs to 
 flush file data to disk, which causes more contention than both {\tt ftruncate} + {\tt fsync}
that  flushes metadata only and  
{\tt fsync}-only that aims to retain a file's durability.
These three settings are commonly I/O behaviors found in today's applications.
Assuming that we make program X probe by calling and measuring {\tt fsync} latency,
it can detect whether applications like program Y are calling {\tt fsync}s
due to the substantial difference with and without contention.
\autoref{fig:raw-trace-ftruncate} shows two raw traces of varying {\tt fsync} latencies
when 1) both programs X and Y are doing with {\tt fsync}-only
and 2) both are doing with {\tt ftruncate} + {\tt fsync}, respectively.
It is evident that the 
latency sensed by program X largely fluctuates
due to the impact of concurrent {\tt fsync}s,
thereby enabling a clear and reliable channel to transmit data.
In all, these qualitative and quantitative observations
indicate the viability of building a covert channel with {\tt fsync} at the persistent storage.

\begin{figure}[t]
	\centering
	\includegraphics[width=0.7\columnwidth]{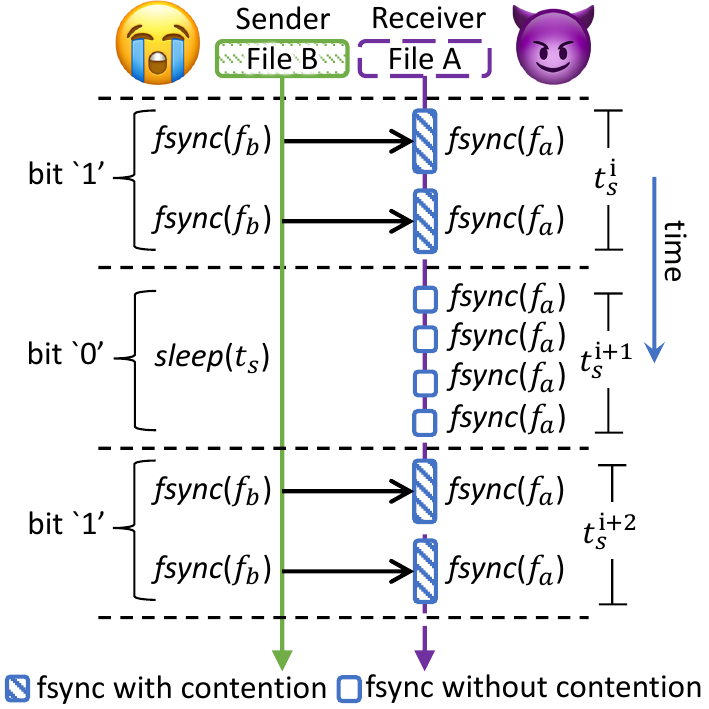}
	\caption{An Illustrative Example of `1' and `0' Transmission Protocol between Sender and Receiver
		with \Cocoa.} %
	\label{fig:algo-send-receive}
\end{figure}

\section{\Cocoa Covert Channel}\label{sec:covert:channel}

Modern 
OS utilizes 
file systems and access control mechanisms 
to isolate files for users and 
store data with secrecy and privacy. %
Cloud vendors enforce further isolation of files for multi-tenants 
to secretly share a physical machine between  containers and VMs. %

In this section, 
we %
focus on leveraging {\tt fsync}s
between two  entities co-located in a storage device
 for secretive and reliable communication, 
regardless of
whether they are %
 running in the same OS, different containers, or VMs.
As both sides call {\tt fsync} to 
interact
through file system and storage device, we name our new covert channel `\Cocoa'.
\Cocoa is timing-based since it transmits information %
 by measuring response time at runtime.

\subsection{Attack Model}\label{sec:covert:model}

We assume that there are two co-located entities in a \Cocoa attack.
The receiver (attacker or sink) and the sender (victim or source) 
concurrently run as user-space programs. %
Both sides %
use files to store data.
We categorize \Cocoa attack cases into three classes
by their isolation environments. 
	{\bf 1) Cross-file}: Sender and receiver access their respective files with 
	exclusive access permissions. 
	These files share the same storage device, residing either
	in the same disk partition mounted with the same file system (e.g., Ext4)
	or different  partitions mounted with the same or different file systems (e.g., Ext4 and XFS).
	{\bf 2) Cross-container}: Sender and receiver stay  in two containers sharing the same  device for their overlay file systems. This
	  is common for hosting multiple containers in a physical machine.
	For example, by default the Docker~\cite{app:Docker}  
	stores all containers' overlay data in one directory {\tt /var/lib/docker/overlay2/}.
	Sender and receiver can also use different directories with different file systems mounted 
	on different partitions.
	{\bf 3) Cross-VM}: Sender and receiver are running in their respective
	 VMs with    independent disk images. Again, 
	their image files need to be co-located in one storage device but can stay in different partitions.

With any channel, both sender and receiver can 
update file size via {\tt ftruncate} + {\tt fsync},
modify their own files via {\tt write} + {\tt fsync}, 
or keep file synchronized to storage device via {\tt fsync}-only (see Section~\ref{sec:viability}).
However, by referring to~\autoref{tab:fsync-latency} and~\autoref{fig:raw-trace-ftruncate}, we 
mainly leverage {\tt fsync}-only to build \Cocoa %
channel,
because it incurs the shortest latency and implies the highest bandwidth for information transmission.

\subsection{Communication Design}\label{sec:covert-design}

To transmit data via any aforementioned channel,
the sender conveys bits by calling {\tt fsync}s on a file   or not.
The receiver continually synchronizes the other file via {\tt fsync} 
and measures the latency of {\tt fsync}
to decide the values of received %
bits.
\autoref{fig:algo-send-receive} shows how sender
and receiver transmit a bit with the protocol provided by 
\Cocoa covert channel.

The sender transmits bits via some purposeful file operations.
As shown by the left part of \autoref{fig:algo-send-receive},
in order to transmit a bit `1',
the sender %
synchronizes a file in order to continuously submit I/O requests 
to disk for a predefined time period {$t_s$} named {\em symbol duration}.
Otherwise, the sender sleeps for $t_s$ to transmit  a bit `0'.
A stream of continuously transmitted bits form a meaningful data frame that %
is %
composed of two parts, i.e., header and payload.
The header consists of a fixed number of bits with a distinct pattern, 
used to accurately distinguish the start (boundary) of a frame.
The payload is a bit stream with a fixed length 
and stores actual data that is useful for the receiver. 

The receiver receives a bit %
by measuring {\tt fsync} latencies to detect whether contention happens %
or not. Before transmission, 
the sender constantly %
 samples the uninterrupted %
  {\tt fsync} latency to profile a threshold for reference. %
In a symbol duration during transmission, the receiver %
checks if {\tt fsync} latency is greater than the profiled threshold.
If so, the bit is `1'; otherwise, the bit is `0'.
The receiver tracks
whether a frame of bits is received by comparing against the header pattern for calibration and synchronization.
In case of %
a match, the receiver extracts payload from the frame.

The threshold to decide `1' or `0'
is set empirically, 
depending on underlying machine's configurations and runtime environmental factors.
For example,
file system and storage device are being used by many %
 programs which %
may cause noise with file operations, especially {\tt fsync}s, to affect \Cocoa. 
We take into account such noise and
select a proper threshold %
in a {heuristic} approach~\cite{security:power:Security-2021}. In short,
we firstly smooth the samples within a symbol duration based on their average. 
Then, we jointly consider polished latencies from numerous continuous symbol durations
to determine the current threshold and periodically update it. 

\begin{figure*}[t]
	\centering
	\begin{subfigure}{\columnwidth}
		\includegraphics[width=\textwidth]{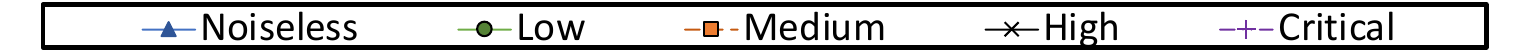}
	\end{subfigure}
	\begin{subfigure}{\textwidth}
		\includegraphics[width=\textwidth]{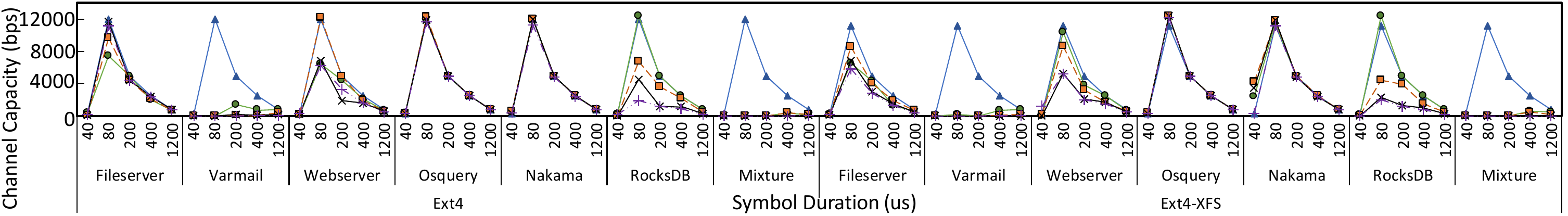}
	\end{subfigure}
	\caption{The Capacity of Cross-file \Cocoa Channel (with and without Noise from Multiple Workloads).}
	\label{fig:bare:channel}
\end{figure*}

\subsection{Performance Evaluation}\label{sec:channel:eval}

\subsubsection{Evaluation Setup}\label{sec:setup}

In order to thoroughly evaluate \Cocoa channel,
we test it 
on a server with Intel Xeon Gold 6342 CPU and 960GB %
SAMSUNG PM883 SATA SSD.
The OS is Ubuntu 21.04 with kernel 5.15.0-48-generic while the compiler
is GCC/G++ 10.3.0. 
We divide SSD space
into few partitions %
and make different file systems on them, 
including Ext4, XFS, and Btrfs.
We use Docker 20.10.14 to manage containers for which the base image is Ubuntu 21.04. %
As for cross-VM testing, the guest OS is %
Ubuntu 21.04 with Ext4 as file system, running
with QEMU and Linux Kernel-based Virtual Machine (KVM) version 4.2.1.
VM disk images are set in the RAW format with virtio enabled. 
We use the default writeback cache policy. 
Each VM is %
with 8GB 
DRAM and two vCPUs.
Sender and receiver programs are respectively
running in two isolated containers (resp. VMs).
In all tests,
both of them are %
normal user-space programs  
and access ordinary files without any  privileged permission.

\begin{table}[t]
	\caption{Noisy Workloads with Different Settings.}
	\label{tab:workloads}
	\resizebox{\columnwidth}{!}{%
		\begin{tabular}{ccccccc}
			\hline
			\multirow{2}{*}{Degree} & \multicolumn{3}{c}{Filebench}~(Number of threads) & \multirow{2}{*}{Osquery} & \multirow{2}{*}{Nakama} & \multirow{2}{*}{RocksDB} \\ \cline{2-4}
			& \multicolumn{1}{l}{Fileserver} & Varmail & Webserver &                    &                              &                          \\ \hline
			Low      & 1& 1& 1                      & Every 60s & 0.1 QPS & 100:0 \\
			Medium   & 2& 2& 2                     & Every 30s & 1 QPS   & 80:20 \\
			High     & 4& 4& 4                      & Every 10s & 10 QPS  & 50:50 \\
			Critical & 8& 8& 8                   & Every 1s  & 100 QPS & 20:80 \\ \hline
		\end{tabular}%
	}
\end{table}

We evaluate three covert channel types, i.e.,
cross-file, cross-container, and cross-VM,
built on different file systems, such as Ext4, XFS, and Btrfs. 
Because of space limitation, we place a part of results and raw traces in~\autoref{sec:app:capacity}. 
In the following contexts, the sender and receiver programs only 
synchronize their respective files using {\tt fsync}, without modifying the files 
in the cross-file and -container channels. 
In the case of the cross-VM channel, 
both the sender and receiver modify the file data before invoking {\tt fsync} on their files.

Besides the noiseless environment, we employ workloads running
alongside \Cocoa to quantitatively assess the robustness of it.
As shown in~\autoref{tab:workloads},
we configure each workload with different settings to
cause environmental noise at varying degrees. %
We adopt 1) Filebench~\cite{benchmark:filebench} to generate Fileserver, Varmail, and Webserver workloads 
with varying threads,
2) an  audit server Osquery~\cite{auditing:Osquery} 
with different auditing frequencies,
3) Tsung~\cite{gameserver:Tsung} that simulates 1000 users concurrently sending
 various requests to a multi-player gaming server Nakama~\cite{gameserver:nakama}
with different Queries Per Second (QPS),
and 4) YCSB~\cite{tool:ycsb} that issues requests
to a prevalent NoSQL database  RocksDB~\cite{db:rocksdb}  
with varying Read/Write ratios.
Furthermore, 
we  conduct experiments by  running all foregoing applications  simultaneously (denoted as Mixture).

\begin{table}[t]
	\centering
	\caption{Raw Bit Error Rate for Cross-file Covert Channel (Ext4, with and without Noise from Fileserver and Varmail).}
	\label{tab:raw-err}
	\resizebox{\columnwidth}{!}{%
		\begin{tabular}{ccccccc}
			\hline
			\multirow{2}{*}{\begin{tabular}[c]{@{}c@{}}Symbol \\ Duration ($\mu$s)\end{tabular}} & \multicolumn{2}{c}{Noiseless} & \multicolumn{2}{c}{Fileserver, 8 Threads} & \multicolumn{2}{c}{Varmail, 8 Threads} \\ \cline{2-7} 
			& \multicolumn{1}{c}{1$\longrightarrow$0} & 0$\longrightarrow$1 & \multicolumn{1}{c}{1$\longrightarrow$0} & 0$\longrightarrow$1 & \multicolumn{1}{c}{1$\longrightarrow$0} & 0$\longrightarrow$1 \\ \hline
			40 & \multicolumn{1}{c}{43.24\%} & 43.24\% & \multicolumn{1}{c}{44.80\%} & 44.81\% & \multicolumn{1}{c}{74.12\%} & 74.12\% \\ %
			50 & \multicolumn{1}{c}{0.40\%} & 0.43\% & \multicolumn{1}{c}{1.84\%} & 1.87\% & \multicolumn{1}{c}{67.38\%} & 67.34\% \\ %
			80 & \multicolumn{1}{c}{0.38\%} & 0.38\% & \multicolumn{1}{c}{1.40\%} & 1.39\% & \multicolumn{1}{c}{66.39\%} & 66.37\% \\ %
			200 & \multicolumn{1}{c}{0.03\%} & 0.03\% & \multicolumn{1}{c}{1.53\%} & 1.43\% & \multicolumn{1}{c}{47.63\%} & 47.62\% \\ %
			400 & \multicolumn{1}{c}{0.00\%} & 0.00\% & \multicolumn{1}{c}{0.34\%} & 0.34\% & \multicolumn{1}{c}{41.01\%} & 41.02\% \\ %
			1200 & \multicolumn{1}{c}{0.00\%} & 0.00\% & \multicolumn{1}{c}{0.89\%} & 0.89\% & \multicolumn{1}{c}{37.73\%} & 37.74\% \\ \hline
		\end{tabular}%
	}
\end{table}

\begin{figure*}[t]
	\centering
	\begin{subfigure}{\columnwidth}
		\includegraphics[width=\textwidth]{new-eval-bar.pdf}
	\end{subfigure}
	\begin{subfigure}{\textwidth}
		\includegraphics[width=\textwidth]{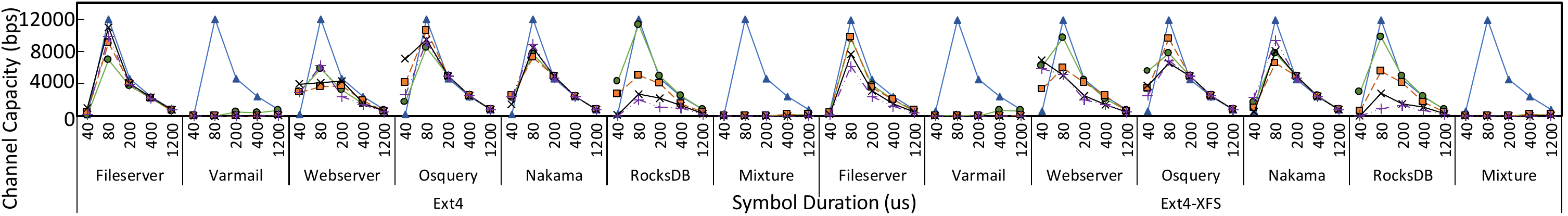}
	\end{subfigure}
	\caption{The Capacity of Cross-container \Cocoa Channel (with and without Noise from Multiple Workloads).}
	\label{fig:docker:channel}
\end{figure*}

\begin{figure*}[t]
	\centering
	\begin{subfigure}{\columnwidth}
		\includegraphics[width=\textwidth]{new-eval-bar.pdf}
	\end{subfigure}
	\begin{subfigure}{\textwidth}
		\includegraphics[width=\textwidth]{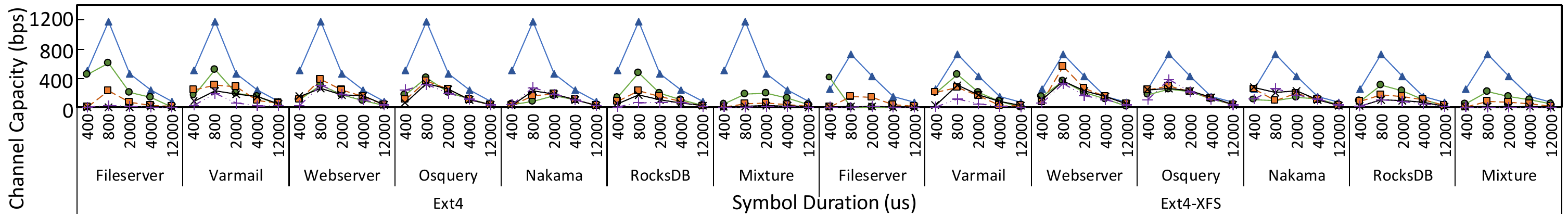}
	\end{subfigure}
	\caption{The Capacity of Cross-VM \Cocoa Channel (with and without Noise from Multiple Workloads).}
	\label{fig:qemu:channel}
\end{figure*}

\subsubsection{Channel Capacity}

We use the channel capacity to measure 
the bandwidth of \Cocoa in bps,
which is a theoretical upper bound for \Cocoa's communication capability.
We regard the \Cocoa channel as a binary symmetric channel 
for measurement, following the methodology in~\cite{security:covert-model}.
The capacity of \Cocoa (denoted as $C$) is determined as follows.
\begin{equation}
	C = B\times{\Big[}1 - p\log_2{(\frac{1}{p})} - (1-p)\log_2{(\frac{1}{1-p})}{\Big]}
\end{equation}
where $B = \frac{1}{t_s}$ is the bandwidth 
and $p$ is the symbol error rate (bit error rate)~\cite{channel:Shannon}
\footnote{Sometimes the error rate might be greater 50\%, 
	and we still regard channel capacity as 0, 
	since channel with high bit error rate cannot transmit data correctly in our cases.}. %
We empirically estimate the bit error rate of \Cocoa channel  every ten frames transmitted, 
each of which carries 8,000 bits.
We also vary the symbol duration $t_s$ 
for each one of three aforementioned covert channels.

{\bf Intra-partition without noise.} %
\autoref{fig:bare:channel},~\autoref{fig:docker:channel}, and~\autoref{fig:qemu:channel} 
show the capacities of all cross-file, -container, and -VM channels,
 respectively.
To conduct a comprehensive study, 
we have performed experiments 
where the sender and receiver are located within the same partition (represented as Ext4 under the X-axis) 
and different partitions (represented as Ext4-XFS under the X-axis) mounted with different file systems. 
The cross-file channel, 
operating within one Ext4 partition, 
achieves a transmission rate of one bit every 50$\mu$s with an error rate of approximately 0.40\% as shown in~\autoref{tab:raw-err}. 
Therefore,
\Cocoa achieves %
20,000bps bandwidth without noise.
The capacity of the cross-container channel, with two containers co-located in a disk partition with Ext4, 
is similar to that of the cross-file channel. 
This can be attributed to the weak isolation provided by the container and overlay file system that
 heavily rely on the underlying file system. 
Meanwhile, the capacity of the cross-VM channel is lower, 
with a maximum bandwidth of about 1200bps. 
The reason is that 
the sender and receiver modify their files before invoking {\tt fsync} in the guest OS 
to trigger stable contention 
at the host OS and underlying storage device.

{\bf Inter-partition without noise.}
In all three channels,
the receiver is placed in a partition managed with Ext4, 
while the sender operates with XFS or Btrfs mounted on another partition of the same SSD. 
The channels using Ext4 and Btrfs (see~\autoref{sec:app:capacity}) 
cannot be established with an acceptable bit error rate ($\le$ 40\%).
This limitation arises because Btrfs does not flush the device cache for {\tt fsync} 
unless there is concrete data modification to the files. 
In contrast, Ext4 and XFS flush the cache on every {\tt fsync}, regardless of file modifications. 
Thus,
the
capacities of cross-file and -container covert channels based on inter-partition with Ext4-XFS 
exhibit similar observations with Ext4 in one partition.
However, due to the influence of the VM hypervisor, 
the cross-container channel  gains a much higher transmission frequency (every 80$\mu$s) 
compared to the cross-VM channel (every 12ms). 
Additionally, the inter-partition %
 cross-VM covert channel  
experiences higher error rates and lower capacities than the intra-partition channel, 
mainly due to the impact of partitions. 
For instance, the capacity with an 800$\mu$s symbol duration of the inter-partition channel 
is 38.1\% lower than that of the intra-partition channel.

\subsubsection{Impact of Noisy Workloads}\label{sec:workload}

Next, we investigate the influence of different workloads on the \Cocoa channel 
in the presence of noise.
Our evaluation reveals that most of the %
workloads introduce noise to all types of \Cocoa channels with varying degrees, thereby
resulting in reduced capacities. 
For example, the bit
error rate shown in~\autoref{tab:raw-err}
for cross-file covert channel with 50$\mu$s duration 
increases to about 1.84\% and 67.34\% when running Fileserver and Varmail with 8 threads, respectively.
As shown in~\autoref{fig:bare:channel}, %
Varmail and  RocksDB have the most significant impact on \Cocoa. 
This is because both of them frequently invoke {\tt fsync}s to ensure the durability of data,
especially for Varmail. 
Fileserver and Webserver also affect the channel capacity, even without {\tt fsync}s, 
particularly with shorter symbol durations. 
This is due to their consistent file operations (e.g., create and delete) and data writes to multiple files, 
leading to the accumulation of dirty pages in the OS's page cache and triggering page write-back. 
Conversely, Osquery and Nakama have limited impact on \Cocoa, 
indicating that \Cocoa is not highly sensitive to auditing activities and network traffic. 
The capacity under mixed applications (Mixture) is similar to that under Varmail, 
because Varmail frequently calls {\tt fsync}s. 
In order to mitigate the impact of such noise,
we can prolong the channel's symbol duration  and 
adopt {\tt fsync} with modifying file data to build a more reliable channel.
As indicated by~\autoref{tab:raw-err}, a longer symbol duration dramatically
decreases the error rate.
The results with cross-container channel are similar to those with the cross-file channel (see \autoref{fig:docker:channel}).

As illustrated by~\autoref{fig:qemu:channel},
the cross-VM covert channel is more susceptible to noise compared to the cross-file and -container channels, 
even for workloads without {\tt fsync}s. 
For example, Fileserver introduces severer noise than Varmail. 
The bit error rate for the cross-VM channel with a symbol duration of 12ms 
increases from around 0.07\% to an average of 33.0\% and 16.4\% 
when running Fileserver and Varmail with 8 threads, respectively. 
We run Fileserver and Varmail in the same disk image under the same guest OS with
the %
sender.
Updating multiple files for Fileserver in the guest OS is actually updating the disk image in the host OS, 
and updated data might not reach the real disk unless calling {\tt fsync}.
Once the sender calls {\tt fsync} in the guest OS, the host OS would synchronize the entire disk image file.
Therefore, 
dirty data written by Fileserver are  flushed alongside into disk with sender's {\tt fsync},
which causes more difficulties with longer {\tt fsync} latency 
for receiver to detect
due to a larger volume of dirty data. 
As to Varmail that frequently calls {\tt fsync} to flush every modified email file, %
each {\tt fsync} is converted to synchronizing the disk image and
does not aggregate a lot of dirty data like Fileserver
when the sender calls {\tt fsync}s.
As a result, the impact of noise caused by Varmail on bit error rate for
\Cocoa is evidently lower than that of Fileserver.

\begin{figure*}[htbp]
	\begin{subfigure}{0.33\textwidth}
		\includegraphics[width=\textwidth,page=1]{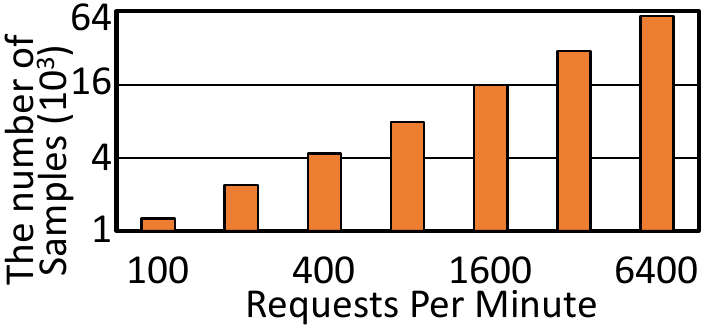}
		\caption{SQLite Insert Requests Estimation.}
		\label{fig:side:insert}
	\end{subfigure}	
	\hfill
	\begin{subfigure}{0.33\textwidth}
		\includegraphics[width=\textwidth,page=2]{new-sqlite.pdf}
		\caption{Raw Trace for Insert with or without Split.}
		\label{fig:side:split}
	\end{subfigure}
	\hfill
	\begin{subfigure}{0.33\textwidth}
		\includegraphics[width=\textwidth,page=3]{new-sqlite.pdf}
		\caption{Estimated Latency Distribution of SQL.}
		\label{fig:side:sql-type}
	\end{subfigure}
	\caption{An Illustration of \Cocoa Side Channel for Databases.}
	\label{fig:side}
\end{figure*}

\section{Side-Channel Attacks with \Cocoa}\label{sec:side:channel}

\subsection{Database Operations Speculation}\label{sec:side:request}

\textbf{Attack Model.} 
In order to perform database operating speculation, 
we consider a scenario where the victim is a database that stores and accesses data using a disk. 
The attacker, who is located on the same disk with the victim, 
operates as an independent process within the same OS, separate container, or VM. 
Particularly, the attacker does not have permission to access the victim's in-memory data or on-disk files. 
To infect the victim's environment, 
the attacker can exploit vulnerabilities via image pollution or social engineering. 
Due to the compactness of the attacker's code, 
it is possible for the attacker to inject the victim gadget into useful applications without perception. %
The attacker has full control over the attacker's program, container, or VM. 
She/he can manipulate containers or VMs to share the same disk with the victim, similar to co-located allocation~\cite{security:cloud:Security-2015,security:clound-vm-allocation:TDSC-2015}.
Additionally, the attacker can control multiple containers or VMs located on different disks and passively wait for a victim to use those disks. 
Once both parties share a device, 
the attacker initiates spying and stealing of sensitive information from the victim using the \Cocoa channel.

\textbf{Attack Design.} 
On a disk, we set up isolated files  in which
a group of them
belong to the victim database for data storage while
and one file is used by the attacker to detect {\tt fsync} latencies. 
We utilize SQLite as the victim database, 
which is widely used and has been exploited for security purposes~\cite{security:SQLite-query:Security-2021,FS:exF2FS:FAST-2022,security:NVLeak:USENIX-2023}. 
We configure SQLite in the {\tt journal\_mode=DELETE} mode and perform various database operations to simulate requests received by the victim.
Simultaneously, the attacker  invokes {\tt fsync} with three objectives. 
Firstly, the attacker aims to monitor the rate of insert and update requests in the victim database (Section~\ref{sec:side:insert-ratio}).
Secondly, the attacker aims to detect internal structural changes in the victim database, such as a node split in the B-Tree used for indexing (Section~\ref{sec:side:split}).
Thirdly, the attacker aims to identify and resemble a sequence of database operations executed by the victim, 
thereby extracting more fine-grained information (Section~\ref{sec:side:sql}).

\subsubsection{Insert/Update Ratio over Time}\label{sec:side:insert-ratio}

Insert and update requests are tightly correlated to the {\tt fsync} latency, 
because when handling such requests,
databases such as SQLite %
utilize redo or undo logging with {\tt fsync}s 
to ensure consistency and durability. 
We adopt Mobibench~\cite{bench:Mobibench,benchmark:mobibench} to repeatedly insert or update 
primary keys and corresponding data into the database, with
 the idle period adjusted for each insert to demonstrate different request frequencies. 
By leveraging the \Cocoa side channel, 
the attacker detects longer {\tt fsync} latencies when the victim database synchronizes
files.
The attacker counts latency samples above a threshold (50$\mu$s in our evaluation) that has been determined through profiling. 
The attacker accordingly
 estimates the victim's insert and update activities and
calculates the rate of insert/update over time.
We repeat this attack for ten times, each in 30 minutes. 

\autoref{fig:side:insert} depicts the relation between 
the number of insert requests per minute for SQLite and 
the number of samples above the threshold, 
using a logarithmic scale for the Y-axis.
The number of samples above the threshold is approximately ten times higher than the number of requests per minute. 
The reason for such an order of magnitude difference 
is twofold.
Firstly, SQLite requires multiple {\tt fsync}s to flush log and data files 
in order to commit a single transaction. 
Secondly, the attacker's {\tt fsync} latency is shorter than that of the victim 
because the victim flushes both data and metadata for database files to complete a database transaction. 
The attacker needs to call {\tt fsync} for multiple times to cover the entire course of a database transaction.
Consequently, the attacker is able to identify when the victim handles an insert or update request and thus calculate the rate of such requests over time. 
In a long run, the attacker can figure out the victim's workload characteristics, such as the insert/update frequency, modified data per request, and peak/non-peak periods in a day.

\subsubsection{B-Tree Split Detection}\label{sec:side:split}

SQLite utilizes an on-disk B-Tree for indexing keys. 
Each B-Tree node has a limited size. 
A fully filled node triggers a split that results in two nodes.
SQLite calls {\tt fsync}s to persist both new nodes to ensure consistency and durability. 
Under a relatively consistent workload, 
a node split %
leads to a longer committing latency for an insert into the B-Tree compared to inserts without node splits. 
Hence, the attacker can detect such structural changes in the SQLite database via \Cocoa side channel.
For example, the raw trace in~\autoref{fig:side:split} demonstrates three different insert requests. 
The latency of middle one experiences longer than latencies of others
due to the occurrence of a node split, 
while the other two  have not involved two nodes to be persisted via {\tt fsync}s.
Additionally, researchers have mentioned 
that sensitive data stored in a B-Tree-based database could be leaked by exploiting node splits~\cite{database:ND2DB}.

The attacker does not have knowledge of when an insert request starts or ends 
but can only determine when the corresponding {\tt fsync} latencies spike and return to normal. 
Consequently, the attacker has introduced an estimated {\tt fsync} latency to speculate whether an insert with a node split has occurred. 
The attacker considers the first sample time with an {\tt fsync} latency greater than a threshold
 as the estimated start timestamp for an insert request. 
The attacker detects the last sample that exceeds the threshold and estimates an end time, 
which is calculated by adding the measured {\tt fsync} latency to the timestamp of the last sample. 
Note that the idle period before the next insert request is expected to be sufficient for the attacker to determine 
if a detected {\tt fsync} belongs to the current insert or not. 
Furthermore, recent studies have shown that typical workloads in commercial environments 
are dominated by inserting small values (less than 1KB or 100B), 
which are unlikely to flush more than a 4KB block~\cite{kv:Facebook-workloads:FAST-2020,kv:LSM-Trie:ATC-2015}. 
Therefore, the impact of a B-Tree node split on two 4KB blocks 
with {\tt fsync}s is realistically substantial. %

The estimated start time and end time for an insert are  exemplified in~\autoref{fig:side:split}.
The estimated {\tt fsync} latency for the victim database exhibits an identical observation to the actual {\tt fsync} latency. 
We insert 400 different primary keys and data into SQLite and cause 49 node splits in all.
We set the threshold at 70$\mu$s to estimate the latency for each insert, 
and empirically classify an insert 
with an estimated {\tt fsync} latency greater than 1000$\mu$s 
as one causing a split. 
The attacker successfully detects 43   node splits, achieving an accuracy of \fpeval{round((43 / 49)*100, 1)}\%, 
and the F1-score is 0.84. 
By leveraging more sophisticated algorithms such as learning-based methods, 
the attacker may have even higher capability to differentiate inserts with node splits from normal inserts. 
This is yet not the focus %
of building \Cocoa side channel in this paper.

\subsubsection{Database Operation Leakage}\label{sec:side:sql}

By utilizing {\tt fsync} latency as a probe, 
we can further categorize various database operations 
to discover and learn about the operations executed by the victim. 
In this study, we create database tables from the NPPES dataset~\cite{dataset:NPPES,security:NVLeak:USENIX-2023} containing two tables.
One table is a {\tt basic} 
that records users' basic information (e.g., their full names), 
using the National Provider Identifiers (NPI) as primary keys. 
The other table named {\tt location} stores users' addresses, including city and state, 
with the NPI serving as a foreign key for the {\tt basic} table.

\begin{lstlisting}[language=SQL, caption=Classified SQL examples., label=code:sql]
	-- I1: Insert 1000 records in 'basic' table
	INSERT INTO basic values (...), (...);
	-- Q1: Count records in 'location' table
	SELECT COUNT(*) FROM location WHERE lower(city)='Anaheim';
	-- U1: Update 1 record in 'basic' table
	UPDATE basic SET name = 'new name' WHERE npi == 20230809;
	-- U2: Update 1000 records in 'location' table
	UPDATE location SET city = 'new city' WHERE npi == 20230809;
\end{lstlisting}

\autoref{code:sql} provides examples of four database operations that the attacker can classify using the \Cocoa side channel. 
{\tt I1} represents the insertion of 1000 records, 
while {\tt Q1} performs queries to count records. 
{\tt U1} updates a single record in the {\tt basic} table. 
Comparatively, {\tt U2} updates 1000 records in the {\tt location} table. 
We repeat the U1 and U2 operations for 100 times and estimate their {\tt fsync} latencies 
using a similar way as presented in Section~\ref{sec:side:split}, with a threshold of 50$\mu$s based on profiling. 
\autoref{fig:side:sql-type} shows the distribution of estimated latencies for each database operation. 
For the {\tt Q1} operations, 
the majority of estimated latencies are close to 0 
because queries do not involve {\tt fsync} to flush data. 
As to update operations ({\tt U1} and {\tt U2}), 
the different numbers of records being modified 
result in varying data volumes for {\tt fsync}s, 
leading to different distributions of estimated latencies. 
Whereas,
the insert operations ({\tt I1}) exhibit latencies that are consistently higher than other operations, 
even though {\tt I1} and {\tt U2} involve the same number of records. 
The reason is that, in contrast to an
update that modifies one node only, some insert may
cause node split that leads to longer latency.

Taking estimated latency for each operation as its feature,
we employ the classic k-nearest neighbors (k-NN) algorithm~\cite{eval:knn} for classification. 
We calculate the Euclidean distances between a newly detected operation and each of known operations. 
Then, we assign a type to the new operation 
based on the majority of the k nearest neighbors. 
We randomly choose 70\% samples for each operation as training set and 
the rest 30\% samples are used as the testing set to evaluate the k-NN algorithm.
In the end, 115 out of 120 operations are correctly classified, resulting in an accuracy of 95.8\% for the \Cocoa side channel. 
The F1-scores for {\tt I1}, {\tt Q1}, {\tt U1}, and {\tt U2} are 0.98, 0.98, 0.93, and 0.95, respectively.

\subsubsection{Comparison}

The way \Cocoa leaks %
 information is not only effectual and portable, 
but also difficult to be perceived. 
For instance, Chen et al.~\cite{security:power:Security-2021} proposed a side channel based on idle power management for CPUs, 
allowing them to spy on a victim's network traffic, such as HTTP traffic load measured in requests per minute. 
However, the information they obtain is more coarse-grained compared to what is achieved with \Cocoa that
 provides a detailed trace of the victim database's operations (insert/query/update). 
Moreover, \Cocoa is built on {\tt fsync}, which is commonly and frequently used by applications on various platforms. 
By contrast, the side channel presented by Chen et al.~\cite{security:power:Security-2021} relies on the prerequisite condition that CPU enters an energy-saving mode and switches back. 
The ND2DB attack~\cite{database:ND2DB} detects B-Tree node splits by measuring the response time of an insert request. 
However, the response time can be easily influenced by user software overhead and network latency. 
In our empirical study, using a threshold of 1300$\mu$s, the ND2DB attack detects only 75.5\% of node splits with an F1-score of 0.65. 
Whereas \Cocoa detects 87.8\% of node splits with an F1-score of 0.84. 
The NVLeak attack~\cite{security:NVLeak:USENIX-2023} can figure out the database operations like \Cocoa 
and achieves an 84\% classification accuracy, which is yet much lower than the 95.8\% accuracy of \Cocoa.
Furthermore, the NVLeak attack only works with Optane memory that Intel has winded down~\cite{news:optane-dead,FS:MAdFS:FAST-2023}. 
Additionally, \Cocoa can distinguish different database operations 
and may jointly make use of 
other techniques %
to inflict more fine-grained leakage with the victim database~\cite{security:database:CCS-2016,security:database:CCS-2018,security:database:SP-2018,security:SQLite-query:Security-2021}.
In all,
the prevalence of {\tt fsync} entitles high efficacy, portability, and unnoticeability to \Cocoa.

\begin{figure*}[t]
	\begin{minipage}[t]{0.33\textwidth}
		\includegraphics[width=\textwidth,page=1]{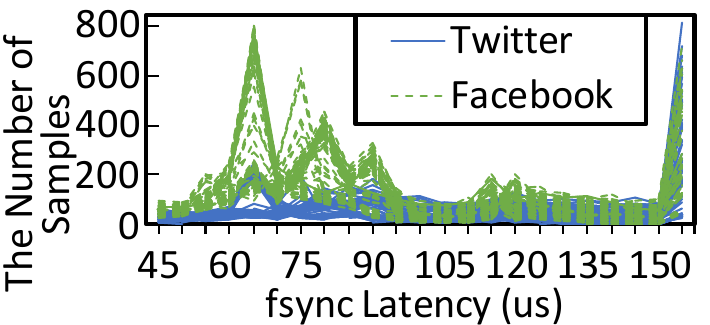}
	\caption{{\tt fsync} Latency Distribution for I/O Traces of Twitter and Facebook in Application Fingerprinting.}
	\label{fig:side:iotrace}
	\end{minipage}
	\hspace{0.002\textwidth}
	\begin{minipage}[t]{0.35\textwidth}
		\includegraphics[width=\textwidth,page=1]{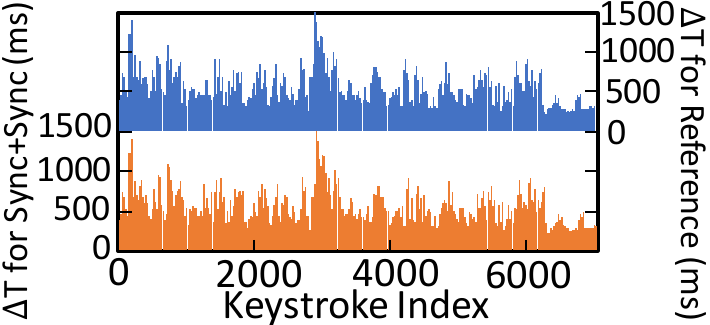}
	\caption{The Inter-keystroke Timings ($\Delta$T) from the Reference Typer (Top, Right Y axis) and the \Cocoa (Bottom, Left Y axis).}
	\label{fig:keystroke:key}
	\end{minipage}
	\hspace{0.002\textwidth}
	\begin{minipage}[t]{0.32\textwidth}
		\includegraphics[width=\textwidth,page=2]{new-keystroke-narrow.pdf}
	\caption{The Time Distribution of the Reference Typers Compared to the Error Distribution of the \Cocoa.}
	\label{fig:keystroke:err}
	\end{minipage}
\end{figure*}

\subsection{Application Information Leakage}

\textbf{Attack Model.} %
We assume a scenario where the victim is an application (e.g., Linux/Android application or 
web browser~\cite{web:Clock,10.1145/3366423.3380124}) that 
accesses files in a disk which
the attacker shares with the victim when
running in the other process, container, or VM.
Again the attacker neither has access permission with
the victim's files nor shares data with the victim.
To leak information from the application, 
the attacker synchronizes her/his file to record {\tt fsync} latencies every 40$\mu$s based on profiling 
and spies on the victim's I/O behaviors.

\textbf{Attack Design.}
To perform information leakage for applications, 
we utilize Mobibench~\cite{bench:Mobibench} to replay I/O traces for victim applications. 
Each I/O trace consists of a series of I/O-related system calls such as {\tt read}, {\tt write}, and {\tt fsync}. 
We execute an I/O trace to simulate the corresponding application's behaviors while the attacker is simultaneously calling {\tt fsync}s.
In practical, 
different applications exhibit varying calling patterns of {\tt fsync}s in terms of frequency and data volume to be flushed. 
These variations result in different {\tt fsync} latencies sensed by the attacker. 
\Cocoa is thus able to distinguish different applications. It also manages to
 fingerprint websites with web browsers under certain conditions.

\textbf{Application Fingerprinting.}  %
We firstly run I/O traces of Twitter and Facebook Android applications, provided by Mobibench, as two victims separately for 100 times.
We record and show the attacker's {\tt fsync} latency distribution  in~\autoref{fig:side:iotrace}. 
It is evident that Twitter and Facebook exhibit distinctly different curves in each of their 100 runs. 
For instance, the number of samples between 50$\mu$s to 100$\mu$s for Facebook is significantly higher than that of Twitter. 
To accurately classify I/O traces for different applications, 
we utilize the latency distribution and Euclidean distances as features and metrics, respectively, for the k-NN algorithm. 
Similar to the k-NN mentioned in Section~\ref{sec:side:sql}, 
we consider the majority of the k nearest neighbors 
for an application trace as the application's type. 
We randomly divide each application trace into 70\% and 30\% samples for training and testing, respectively. 
In the evaluation, we have 60 test cases (100 $\times$ 30\% for each application), and \Cocoa correctly categorizes all of them.

\begin{table}[t]
    \caption{Classification Accuracy of \Cocoa for Websites.}
    \label{tab:website}
    \resizebox{\columnwidth}{!}{%
    \begin{tabular}{ccccccccc}
    \cline{1-4} \cline{6-9}
    Website &
      \begin{tabular}[c]{@{}c@{}}Average\\ \# {\tt fsync}\end{tabular} &
      Accuracy &
      F1-score &
       &
      Website &
      \begin{tabular}[c]{@{}c@{}}Average\\ \# {\tt fsync}\end{tabular} &
      Accuracy &
      F1-score \\ \cline{1-4} \cline{6-9} 
      360.cn        & 10.6  & 3.3\%   & 0.04 &  & imdb.com      & 16.1 & 13.3\% & 0.19 \\
      adobe.com     & 11.5 & 0.0\%  & 0.00 & & jd.com        & 14.9  & 56.7\%  & 0.65 \\
      amazon.com    & 16.2  & 13.3\%  & 0.08 & & live.com      & 10.6  & 13.3\%  & 0.14 \\
      apple.com     & 11.5 & 16.7\% & 0.17 &  & microsoft.com & 12.1 & 3.3\%  & 0.04 \\
      baidu.com     & 14.4  & 6.7\%   & 0.03 & & {\bf  qq.com}      & {\bf 264.6} & {\bf 100.0\% }& {\bf 1.00} \\
      bing.com      & 15.0 & 6.7\%  & 0.08 & & {\bf sina.com.cn} &{\bf 40.8}  & {\bf 96.7\% } &{\bf 0.98}  \\
      booking.com   & 15.9 & 0.0\%  & 0.00 & & sohu.com      & 14.4  & 46.7\%  & 0.44 \\
      cnn.com       & 15.2 & 6.7\%  & 0.10 & & taobao.com    & 10.6  & 23.3\%  & 0.16  \\
      detik.com     & 10.1 & 10.0\% & 0.13 & & tmall.com     & 11.4  & 3.3\%   & 0.03 \\
      github.com    & 12.1 & 13.3\% & 0.18 & & yahoo.co.jp   & 11.4 & 6.7\%  & 0.06 \\ \cline{1-4} \cline{6-9} 
    \end{tabular}%
    }
\end{table}

\textbf{Website Fingerprinting.} %
Next, to check if \Cocoa can distinguish different websites based on their {\tt fsync} usage patterns, 
we create an evaluation dataset as follows. 
Firstly, we randomly select 20 websites from the Alexa Top 100 websites. 
Then, with the Chrome web browser (Version 113 with default settings)~\cite{chrome:chrome}
running on a computer installed with Ubuntu 22.04, 
we visit the front page of each website as the victim. 
Simultaneously, we capture the I/O trace of each website using strace~\cite{linux:strace} for 5 seconds, 
which is sufficient for loading a webpage. %
We repeat this procedure for 100 times with each website and overall collect 2000 I/O traces.

With these I/O traces, we record the attacker's {\tt fsync} latency distribution when replaying each I/O trace 
and classify them using the k-NN algorithm with the same setup as application fingerprinting. 
\autoref{tab:website} shows the classification accuracy and F1-score for each website, 
as well as the average number of {\tt fsync}s invoked by each website.
Most websites do not commonly use {\tt fsync} and hence have similar {\tt fsync} usage patterns. %
As a result, it is challenging for \Cocoa to distinguish them from each other. 
However, some websites, such as {qq.com} and {sina.com.cn}, 
invoke {\tt fsync} more frequently and exhibit different I/O behaviors. 
Therefore, \Cocoa recognizes these websites at high accuracies of, for example, 100\% and 96.7\% for {qq.com} and {sina.com.cn}, respectively. 
Our analysis shows that,
 to persistently store data,
both websites use the Indexed Database~\cite{w3c:indexeddb} %
that most of the browsers provides
~\cite{safari:indexeddb,firefox:indexeddb,chrome:indexeddb,Chromium:indexeddb}.
As mentioned, database relies on  {\tt fsync} to ensure data consistency and durability.
This explains why \Cocoa successfully fingerprints   websites that frequently synchronizes data with the Indexed Database.

\textbf{Comparison.} Distinguishing (fingerprinting) applications and websites
enables an attacker to infer 
which application a victim is using or what website a browser is displaying, 
causing serious breach of user privacy~\cite{web:defense,web:Touching,web:k-finger,security:browser-storage:ACSAC-2016}.
\Cocoa shares similarities with the attack proposed by Kim et al.~\cite{security:browser-storage:ACSAC-2016}, 
as they both exploit storage to fingerprint victims. 
However, they differ in the explored observations.  
Kim et al.~\cite{security:browser-storage:ACSAC-2016} use the disk space quota demanded by a web browser for each website's temporary storage, 
while \Cocoa is based on different {\tt fsync} usage patterns. 
Website fingerprinting attack conducted by Kim et al.~\cite{security:browser-storage:ACSAC-2016} achieves a 97.3\% inference accuracy, 
whereas \Cocoa's accuracy varies depending on the characteristics of websites. 
Additionally, the side channel studied by Kim et al.~\cite{security:browser-storage:ACSAC-2016} only applies to browsers, 
while \Cocoa is widely applicable to applications that call {\tt fsync}.

\subsection{Keystroke Attack}

\textbf{Attack Model.} 
We assume that the victim is entering user input either locally or remotely through a network connection. 
Each keystroke is then transmitted to a service program, which stores the input on an SSD similar to the attack studied by Liu et al.~\cite{security:pmem:USENIX-2023}. 
For every keystroke typed by the victim, 
the service program {auto-commits} %
the user input by storing it in a file with an {\tt fsync}, 
in order to persistently track the user's latest input. 
The attacker and  the service program are co-located to be sharing the same disk.  
However, due to OS-level isolation, any direct communication between them is not possible. 
Also the attacker has no access permission  to the victim's and the service program's files or share any data with the latter two.

\textbf{Attack Design.}
To conduct a keystroke attack, we utilize the Keystroke100 dataset~\cite{dataset:keystroke}. 
It contains inter-keystroke latencies from 100 different typers 
who entered the same eight-letter password, `try4-mbs', ten times each, 
resulting in a total of 7,000 inter-keystroke timings. 
In our attack scenario, the victim sends %
keystrokes to the service program 
with the corresponding delays of inter-keystroke timings prerecorded in~\cite{dataset:keystroke}. 
The service program receives the user input and stores it   using {\tt fsync}. 
In the meantime, the attacker performs the \Cocoa attack by continuously invoking {\tt fsync} 
to infer whether a user input is stored by measuring the attacker's {\tt fsync} latency. 
Given a stored user input,
the attacker can detect an increased {\tt fsync} latency. %
Conversely, if the {\tt fsync} latency remains low, the attacker  deduces with a high probability 
that no user input has been sent to the service program. 
In our evaluation, we set a threshold of 54$\mu$s through profiling to distinguish if a user input is received and stored.

To assess the accuracy of \Cocoa, 
we calculate the timing difference between the ground-truth latencies from the prerecorded dataset and detected ones. 
\autoref{fig:keystroke:key} illustrates a back-to-back comparison of the inter-keystroke timings between the reference typers and   \Cocoa attack. 
The difference %
is negligible, 
as  \Cocoa successfully detects keystrokes with an accuracy of 99.2\%. 
 \Cocoa may misjudge keystrokes due to noise and less intense contention. 
\autoref{fig:keystroke:err} presents the error distribution of   \Cocoa side channel 
in comparison to the timing distributions of the ground truth. 
On average, the error in received timings for   \Cocoa side channel is 2.5ms, with 98.1\% of errors being less than 10ms.
This further justifies the capability of \Cocoa since an inter-keystroke latency
is generally no less than 100ms.

\textbf{Comparison.} %
Inter-keystroke timing has been widely considered in software-based side-channel attacks. 
It allows attackers to reveal sensitive information through simple statistical techniques using keystroke timings~\cite{keystroke:ssh,security:power:Security-2021}.
For instance, Song et al.~\cite{keystroke:ssh} demonstrate that attackers can uncover information about the keys typed 
by analyzing users' typing patterns to recover passwords entered during SSH connections. 
Unlike Gruss et al.~\cite{security:pagecache:CCS-2019} that detect a keystroke when the OS's page cache loads related pages upon a key input issued by the user,
\Cocoa functions at the persistent storage   with a hypothesis that {\tt fsync} operations
are needed to store user input keys. 
\Cocoa is hence infeasible for applications 
that do not auto-commit and invoke {\tt fsync} on the arrival of user input.
With precise keystroke timings, \Cocoa can jointly work with
 techniques like machine learning %
to guess passwords or infer written characters~\cite{keystroke:ssh,security:power:Security-2021,security:pmem:USENIX-2023,keystroke:machine-learning,keystroke:multi-user}.

Recently Chen et al.~\cite{security:power:Security-2021} performed a keystroke attack 
by measuring the uncore power status, 
as the network traffic and encryption stack of SSH connections   affect a system's uncore power. 
They achieved an F1-score of 0.93 with an error rate of 4.9\%. 
Later
Liu et al.~\cite{security:pmem:USENIX-2023} did a similar keystroke attack on Intel Optane persistent memory (PMEM). 
They persistently store the user input to a key-value store on PMEM 
after every keystroke typed by the victim 
and 
detect inter-keystroke timings by probing the Read-Modify-Write (RMW) buffer of PMEM to distinguish RMW hits from misses. 
They achieved an F1-score of 0.99 with an error rate of 1.04\%. 
Comparatively, \Cocoa achieves not only an F1-score of 0.996, but also with a lower error rate of 0.81\%. 

\section{Discussions}\label{sec:discussion}

In this section, we further study
the impact of fast commit in Ext4, 
DoS attack, and
the defense mechanisms about \Cocoa.
In~\autoref{sec:app:discussion},
we also explore the cross-disk \Cocoa channel, 
the channel on other platforms and NVMe SSD, 
and the impact of Pass-through Disk in VM. %

\begin{table}[t]
    \centering
	\caption{Raw Bit Error Rate with and without Fast Commit.}
    \label{tab:err-fast-commit}
    \resizebox{\columnwidth}{!}{%
    \begin{tabular}{ccccc}
    \hline
    \multirow{2}{*}{\begin{tabular}[c]{@{}c@{}}Symbol\\ Duration ($\mu$s)\end{tabular}} & 
    \multicolumn{2}{c}{\begin{tabular}[c]{@{}c@{}}Sender: {\tt ftruncate} + {\tt fsync}\\ Receiver: {\tt fsync}-only\end{tabular}} & \multicolumn{2}{c}{\begin{tabular}[c]{@{}c@{}}Sender: {\tt ftruncate} + {\tt fsync}\\ Receiver: {\tt ftruncate} + {\tt fsync}\end{tabular}} \\ \cline{2-5} 
     & \multicolumn{1}{c}{Fast Commit} & Normal Commit & \multicolumn{1}{c}{Fast Commit} & Normal Commit \\ \hline
    200 & \multicolumn{1}{c}{0.09\%} & 0.15\% & \multicolumn{1}{c}{15.93\%} & 23.75\% \\ %
    400 & \multicolumn{1}{c}{0.06\%} & 0.07\% & \multicolumn{1}{c}{2.65\%} & 2.59\% \\ %
    800 & \multicolumn{1}{c}{0.01\%} & 0.17\% & \multicolumn{1}{c}{0.16\%} & 0.12\% \\ %
    1600 & \multicolumn{1}{c}{0.06\%} & 0.03\% & \multicolumn{1}{c}{0.03\%} & 0.10\% \\ %
    2400 & \multicolumn{1}{c}{0.04\%} & 0.10\% & \multicolumn{1}{c}{0.14\%} & 0.06\% \\ \hline
    \end{tabular}%
    }
\end{table}

\textbf{The Impact of Fast Commit in Ext4.}
Fast commit is a new feature introduced in Ext4 to eliminate unrelated data when invoking {\tt fsync} 
so as to reduce contention for {\tt fsync}s~\cite{journal:iJournaling:ATC-2017,FS:fast-commit-Ext4}.
We test if \Cocoa still functions with fast commit or not.
To
trigger Ext4 journaling every {\tt fsync}, 
sender utilizes {\tt ftruncate} to change file size randomly 
and invokes {\tt fsync} to enable \Cocoa channel, i.e., {\tt ftruncate}+{\tt fsync}. 
Receiver works with {\tt ftruncate}+{\tt fsync} or {\tt fsync}-only. 
\autoref{tab:err-fast-commit} presents bit error rates for \Cocoa channel with and without fast commit. 
Evidently
fast commit does not affect \Cocoa, since 
at the same symbol duration %
\Cocoa yields identical bit error rate despite the use of fast commit.

\begin{figure}[tbp]
    \centering
    \includegraphics[width=\columnwidth]{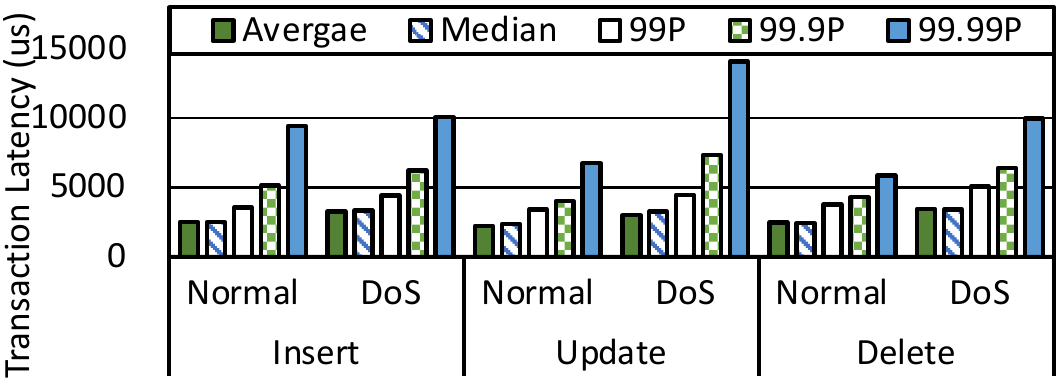}
    \caption{Transaction Average and Tail Latencies for SQLite in VM with and without DoS Attack.}
	\label{fig:dos}
\end{figure}

\textbf{DoS Attack.}
The interference between {\tt fsync}s
 impairs performance and implies 
the potential of DoS attack for \Cocoa.
As VMs isolate programs with hypervisor handling I/Os for them,
the difficulty of launching DoS attacks is   higher than
doing so with programs co-located in the same OS. 
We
run two VMs with independent disk images. %
The attacker in malicious VM continually %
writes its file and invokes {\tt fsync}s. 
We utilize Mobibench to issue 
Insert/Update/Delete workloads with SQLite for 100,000 times 
in  the victim VM.
As shown in~\autoref{fig:dos},
the throughput of SQLite degrades by
\fpeval{round((1 - 307.7 / 397.1)*100, 1)}\%,
\fpeval{round((1 - 333.3 / 449.6)*100, 1)}\%, and
\fpeval{round((1 - 290.17 / 405.09)*100, 1)}\% with
Insert, Update, and Deletion requests, respectively.
The 99.9P (99.9th percentile) tail latency increases %
  by 
\fpeval{round((6193.93 / 5126.32 - 1)*100, 1)}\%,
\fpeval{round((7329.53 / 4022.95  - 1)*100, 1)}\%, and 
\fpeval{round((6372.83 / 4314.93 - 1)*100, 1)}\%, respectively.
These justify \Cocoa's capability in making DoS attacks.

\textbf{Defense Mechanisms.}
It is impracticable  for applications to avoid using {\tt fsync}s. %
One straightforward way to defend against \Cocoa attacks
is to prevent applications %
from being co-located with  other users/applications
in the same machine, especially on the same storage device. %
Yet such hardware isolation %
increases the operating cost
with extra devices or even machines 
and causes waste of disk space.

Using network-based distributed file systems (e.g., Ceph~\cite{filesystem:Ceph}), 
to host applications remotely helps to mitigate  the effect of \Cocoa, as
 {\tt fsync} operations become more sophisticated.
Firstly, {\tt fsync} latency is not only influenced 
by the storage devices, %
but also   affected by the network.
Secondly, distributed file systems often deploy replica %
in different physical machines 
to provide data %
consistency and durability, 
which increases difficulty of generating contention for adversaries to detect.
However, 
this may not be feasible for
some applications
that
are sensitive to network delay and 
need local storage to provide high  performance.
Such applications can consider utilizing a background thread that randomly calls {\tt fsync}s  
to create noise for \Cocoa,
since \Cocoa is sensitive to noise. %
However, to blur {\tt fsync} latencies by introducing extra {\tt fsync}s
may incur performance penalty.

\section{Conclusion}\label{sec:conclusion}

In this paper, we build 
a new \Cocoa side channel
that entails a practical concern   
for programs that are co-located in persistent storage
and call {\tt fsync}s for durability. %
For example,
\Cocoa on Ext4
and an ordinary SSD establishes a covert channel with 
20,000bps bandwidth at about 0.40\% error rate.
\Cocoa is made effectual by the contention between concurrent {\tt fsync}s %
on sharing file system's structures and storage device's hardware resources.
Extensive experiments show that,
leveraging \Cocoa we manage to launch concrete attacks, such as
precisely detecting operations of victim database,
 distinguishing applications, and fingerprinting websites.
We have verified the viability of \Cocoa on various platforms including Linux, Windows, and MacOS,
and responsibly disclosed \Cocoa to the security teams of Linux, Microsoft, and Apple. 
We hope \Cocoa could encourage researchers to further study side-channel information 
leakage at the persistent storage.

\section*{Acknowledgement}

We sincerely thank our shepherd, all reviewers, and the TPC for their valuable comments and suggestions.
We also thank Zhice Yang, Sudipta Chattopadhyay, Fu Song, Guangke Chen, and Lixiang Lian for their helpful discussions.
This work was jointly supported by 
Natural Science Foundation of Shanghai under Grants No. 22ZR1442000 and 23ZR1442300, and ShanghaiTech Startup Funding.

\bibliographystyle{plain}
\bibliography{security}

\begin{thebibliography}{10}

\bibitem{gpu:GPUCC}
Jaeguk Ahn, Jiho Kim, Hans Kasan, Leila Delshadtehrani, Wonjun Song, Ajay
  Joshi, and John Kim.
\newblock Network-on-chip microarchitecture-based covert channel in {GPUs}.
\newblock In {\em MICRO-54: 54th Annual IEEE/ACM International Symposium on
  Microarchitecture}, MICRO '21, page 565–577, New York, NY, USA, 2021.
  Association for Computing Machinery.

\bibitem{w3c:indexeddb}
Ali Alabbas and Joshua Bell.
\newblock Indexed database api 3.0.
\newblock \url{https://www.chromium.org/developers/design-documents/indexeddb},
  2023.

\bibitem{news:optane-dead}
Paul Alcorn.
\newblock Intel kills {Optane} memory business, pays \textdollar559 million
  inventory write-off.
\newblock
  \url{https://www.tomshardware.com/news/intel-kills-optane-memory-business-for-good},
  August 2022.

\bibitem{eval:knn}
N.~S. Altman.
\newblock An introduction to kernel and nearest-neighbor nonparametric
  regression.
\newblock {\em The American Statistician}, 46(3):175--185, 1992.

\bibitem{linux:blktrace}
Jens Axboe, Alan~D. Brunelle, and Nathan Scott.
\newblock blktrace(8) - linux man page.
\newblock \url{https://linux.die.net/man/8/blktrace}, 2006.

\bibitem{security:DUPEFS:USENIX-2023}
Andrei Bacs, Saidgani Musaev, Kaveh Razavi, Cristiano Giuffrida, and Herbert
  Bos.
\newblock {DUPEFS}: Leaking data over the network with filesystem deduplication
  side channels.
\newblock In {\em 20th USENIX Conference on File and Storage Technologies (FAST
  22)}, pages 281--296, Santa Clara, CA, February 2022. USENIX Association.

\bibitem{chrome:indexeddb}
Kayce Basques.
\newblock View and change {IndexedDB} data.
\newblock \url{https://developer.chrome.com/docs/devtools/storage/indexeddb/},
  2023.

\bibitem{bio:multi-queue}
Matias Bj\o{}rling, Jens Axboe, David Nellans, and Philippe Bonnet.
\newblock Linux block {IO}: Introducing multi-queue {SSD} access on multi-core
  systems.
\newblock In {\em Proceedings of the 6th International Systems and Storage
  Conference}, SYSTOR '13, New York, NY, USA, 2013. Association for Computing
  Machinery.

\bibitem{web:Touching}
Xiang Cai, Xin~Cheng Zhang, Brijesh Joshi, and Rob Johnson.
\newblock Touching from a distance: Website fingerprinting attacks and
  defenses.
\newblock In {\em Proceedings of the 2012 ACM Conference on Computer and
  Communications Security}, CCS '12, page 605–616, New York, NY, USA, 2012.
  Association for Computing Machinery.

\bibitem{kv:Facebook-workloads:FAST-2020}
Zhichao Cao, Siying Dong, Sagar Vemuri, and David H.~C. Du.
\newblock Characterizing, modeling, and benchmarking {RocksDB} key-value
  workloads at {Facebook}.
\newblock In {\em Proceedings of the 18th USENIX Conference on File and Storage
  Technologies}, FAST'20, page 209–224, USA, 2020. USENIX Association.

\bibitem{dataset:NPPES}
{Centers for Medicare \& Medicaid Services}.
\newblock {NPPES} dataset.
\newblock \url{https://download.cms.gov/nppes/NPI_Files.html}, December 2019.

\bibitem{filesystem:Ceph}
{Ceph Open Source Community}.
\newblock Ceph file system --- ceph documentation.
\newblock \url{https://docs.ceph.com/en/quincy/cephfs/index.html}, 2016.

\bibitem{FS:OPTR:ATC-2019}
Yun-Sheng Chang and Ren-Shuo Liu.
\newblock {OPTR}: Order-preserving translation and recovery design for {SSDs}
  with a standard block device interface.
\newblock In {\em Proceedings of the 2019 USENIX Conference on Usenix Annual
  Technical Conference}, USENIX ATC '19, page 1009–1023, USA, 2019. USENIX
  Association.

\bibitem{linux:strace}
Vitaly Chaykovsky.
\newblock linux syscall tracer.
\newblock \url{https://strace.io/}, 2023.

\bibitem{security:power:Security-2021}
Paizhuo Chen, Lei Li, and Zhice Yang.
\newblock {Cross-VM} and {Cross-Processor} covert channels exploiting processor
  idle power management.
\newblock In {\em 30th USENIX Security Symposium (USENIX Security 21)}, pages
  733--750, Virtual Event, August 2021. USENIX Association.

\bibitem{app:SQLite}
SQLite Consortium.
\newblock {SQLite}.
\newblock \url{https://www.sqlite.org/index.html}.
\newblock Accessed: 2022-8-31.

\bibitem{tool:ycsb}
Brian~F. Cooper, Adam Silberstein, Erwin Tam, Raghu Ramakrishnan, and Russell
  Sears.
\newblock Benchmarking cloud serving systems with {YCSB}.
\newblock In {\em Proceedings of the 1st ACM Symposium on Cloud Computing},
  SoCC '10, page 143–154, New York, NY, USA, 2010. Association for Computing
  Machinery.

\bibitem{app:Docker}
{Docker Inc.}
\newblock Docker overview.
\newblock \url{https://docs.docker.com/get-started/overview/}, February 2023.

\bibitem{keystroke:machine-learning}
Kwesi Elliot, Jonathan Graham, Yusef Yassin, Trenton Ward, John Caldwell, and
  Tawab Attie.
\newblock A comparison of machine learning algorithms in keystroke dynamics.
\newblock In {\em 2019 International Conference on Computational Science and
  Computational Intelligence (CSCI)}, pages 127--132, 2019.

\bibitem{bench:Mobibench}
ESOS-Lab.
\newblock Mobile benchmark tool (mobibench).
\newblock \url{https://github.com/ESOS-Lab/Mobibench}, May 2020.

\bibitem{branch:JumpOverASLR}
Dmitry Evtyushkin, Dmitry Ponomarev, and Nael Abu-Ghazaleh.
\newblock Jump over aslr: Attacking branch predictors to bypass {ASLR}.
\newblock In {\em The 49th Annual IEEE/ACM International Symposium on
  Microarchitecture}, MICRO-49. IEEE Press, 2016.

\bibitem{branch:BranchScope}
Dmitry Evtyushkin, Ryan Riley, Nael~CSE Abu-Ghazaleh, ECE, and Dmitry
  Ponomarev.
\newblock {BranchScope}: A new side-channel attack on directional branch
  predictor.
\newblock In {\em Proceedings of the Twenty-Third International Conference on
  Architectural Support for Programming Languages and Operating Systems},
  ASPLOS '18, page 693–707, New York, NY, USA, 2018. Association for
  Computing Machinery.

\bibitem{safari:indexeddb}
David Fahlander.
\newblock {IndexedDB} on {Safari}.
\newblock \url{https://dexie.org/docs/IndexedDB-on-Safari}, 2023.

\bibitem{benchmark:filebench}
Filebench.
\newblock Filebench: File system and storage benchmark that uses a custom
  language to generate a large variety of workloads.
\newblock \url{https://github.com/filebench/filebench}, February 2020.

\bibitem{database:ND2DB}
Ariel Futoransky, Dami\'{a}n Saura, and Ariel Waissbein.
\newblock The {ND2DB} attack: Database content extraction using timing attacks
  on the indexing algorithms.
\newblock In {\em Proceedings of the First USENIX Workshop on Offensive
  Technologies}, WOOT '07, USA, 2007. USENIX Association.

\bibitem{FS:soft-updates}
Gregory~R. Ganger, Marshall~Kirk McKusick, Craig A.~N. Soules, and Yale~N.
  Patt.
\newblock Soft updates: A solution to the metadata update problem in file
  systems.
\newblock {\em ACM Trans. Comput. Syst.}, 18(2):127–153, may 2000.

\bibitem{security:cgroups-sync:CCS-2019}
Xing Gao, Zhongshu Gu, Zhengfa Li, Hani Jamjoom, and Cong Wang.
\newblock Houdini's escape: Breaking the resource rein of {Linux} control
  groups.
\newblock In {\em Proceedings of the 2019 ACM SIGSAC Conference on Computer and
  Communications Security}, CCS '19, page 1073–1086, New York, NY, USA, 2019.
  Association for Computing Machinery.

\bibitem{security:covert-model}
Steven Gianvecchio, Haining Wang, Duminda Wijesekera, and Sushil Jajodia.
\newblock Model-based covert timing channels: Automated modeling and evasion.
\newblock In {\em Proceedings of the 11th International Symposium on Recent
  Advances in Intrusion Detection}, RAID '08, page 211–230, Berlin,
  Heidelberg, 2008. Springer-Verlag.

\bibitem{chrome:chrome}
Google.
\newblock {Google} {Chrome} - download the fast, secure browser from {Google}.
\newblock \url{http://google.com/chrome}, 2023.

\bibitem{tlb:TLBleed}
Ben Gras, Kaveh Razavi, Herbert Bos, and Cristiano Giuffrida.
\newblock Translation leak-aside buffer: Defeating cache side-channel
  protections with {TLB} attacks.
\newblock In {\em Proceedings of the 27th USENIX Conference on Security
  Symposium}, SEC'18, page 955–972, USA, 2018. USENIX Association.

\bibitem{security:database:CCS-2018}
Paul Grubbs, Marie-Sarah Lacharite, Brice Minaud, and Kenneth~G. Paterson.
\newblock Pump up the volume: Practical database reconstruction from volume
  leakage on range queries.
\newblock In {\em Proceedings of the 2018 ACM SIGSAC Conference on Computer and
  Communications Security}, CCS '18, page 315–331, New York, NY, USA, 2018.
  Association for Computing Machinery.

\bibitem{security:pagecache:CCS-2019}
Daniel Gruss, Erik Kraft, Trishita Tiwari, Michael Schwarz, Ari Trachtenberg,
  Jason Hennessey, Alex Ionescu, and Anders Fogh.
\newblock Page cache attacks.
\newblock In {\em Proceedings of the 2019 ACM SIGSAC Conference on Computer and
  Communications Security}, CCS '19, page 167–180, New York, NY, USA, 2019.
  Association for Computing Machinery.

\bibitem{security:prefetch+reload:SP-2022}
Yanan Guo, Andrew Zigerelli, Youtao Zhang, and Jun Yang.
\newblock Adversarial prefetch: New cross-core cache side channel attacks.
\newblock In {\em 2022 IEEE Symposium on Security and Privacy (SP)}, pages
  1458--1473, 2022.

\bibitem{security:clound-vm-allocation:TDSC-2015}
Yi~Han, Jeffrey Chan, Tansu Alpcan, and Christopher Leckie.
\newblock Using virtual machine allocation policies to defend against
  co-resident attacks in cloud computing.
\newblock {\em IEEE Trans. Dependable Secur. Comput.}, 14(1):95–108, jan
  2017.

\bibitem{web:k-finger}
Jamie Hayes and George Danezis.
\newblock k-fingerprinting: A robust scalable website fingerprinting technique.
\newblock In {\em 25th USENIX Security Symposium (USENIX Security 16)}, pages
  1187--1203, Austin, TX, August 2016. USENIX Association.

\bibitem{gameserver:nakama}
{Heroic Labs}.
\newblock Nakama server.
\newblock \url{https://heroiclabs.com/docs/nakama/getting-started/index.html},
  2023.

\bibitem{benchmark:mobibench}
Sooman Jeong, Kisung Lee, Jungwoo Hwang, Seongjin Lee, and Youjip Won.
\newblock {AndroStep}: Android storage performance analysis tool.
\newblock In Stefan Wagner and Horst Lichter, editors, {\em Software
  Engineering 2013 - Workshopband}, pages 327--340, Bonn, 2013. Gesellschaft
  für Informatik e.V.

\bibitem{security:database:CCS-2016}
Georgios Kellaris, George Kollios, Kobbi Nissim, and Adam O'Neill.
\newblock Generic attacks on secure outsourced databases.
\newblock In {\em Proceedings of the 2016 ACM SIGSAC Conference on Computer and
  Communications Security}, CCS '16, page 1329–1340, New York, NY, USA, 2016.
  Association for Computing Machinery.

\bibitem{power:powerchannels}
S.~Karen Khatamifard, Longfei Wang, Amitabh Das, Selcuk Kose, and Ulya~R.
  Karpuzcu.
\newblock {POWERT} channels: A novel class of covert communication exploiting
  power management vulnerabilities.
\newblock In {\em 2019 IEEE International Symposium on High Performance
  Computer Architecture (HPCA)}, pages 291--303, 2019.

\bibitem{security:browser-storage:ACSAC-2016}
Hyungsub Kim, Sangho Lee, and Jong Kim.
\newblock Inferring browser activity and status through remote monitoring of
  storage usage.
\newblock In {\em Proceedings of the 32nd Annual Conference on Computer
  Security Applications}, ACSAC '16, page 410–421, New York, NY, USA, 2016.
  Association for Computing Machinery.

\bibitem{journal:Z-journal:ATC-2021}
Jongseok Kim, Cassiano Campes, Joo-Young Hwang, Jinkyu Jeong, and Euiseong Seo.
\newblock {Z-Journal}: Scalable {Per-Core} journaling.
\newblock In {\em 2021 USENIX Annual Technical Conference (USENIX ATC 21)},
  pages 893--906. USENIX Association, July 2021.

\bibitem{cpu:NetCAT}
Michael Kurth, Ben Gras, Dennis Andriesse, Cristiano Giuffrida, Herbert Bos,
  and Kaveh Razavi.
\newblock {NetCAT}: Practical cache attacks from the network.
\newblock In {\em 2020 IEEE Symposium on Security and Privacy (SP)}, pages
  20--38, 2020.

\bibitem{security:database:SP-2018}
Marie-Sarah Lacharit\'e, Brice Minaud, and Kenneth~G. Paterson.
\newblock Improved reconstruction attacks on encrypted data using range query
  leakage.
\newblock In {\em 2018 IEEE Symposium on Security and Privacy (SP)}, pages
  297--314, 2018.

\bibitem{FS:WALDIO:ATC-2015}
Wongun Lee, Keonwoo Lee, Hankeun Sun, Wook-Hee Kim, Beomseok Nam, and Youjip
  Won.
\newblock {WALDIO}: Eliminating the filesystem journaling in resolving the
  journaling of journal anomaly.
\newblock In {\em Proceedings of the 2015 USENIX Conference on Usenix Annual
  Technical Conference}, USENIX ATC '15, page 235–247, USA, 2015. USENIX
  Association.

\bibitem{security:Prime+Probe:SP-2015}
Fangfei Liu, Yuval Yarom, Qian Ge, Gernot Heiser, and Ruby~B. Lee.
\newblock Last-level cache side-channel attacks are practical.
\newblock In {\em 2015 IEEE Symposium on Security and Privacy}, pages 605--622,
  2015.

\bibitem{security:pmem:USENIX-2023}
Sihang Liu, Suraaj Kanniwadi, Martin Schwarzl, Andreas Kogler, Daniel Gruss,
  and Samira Khan.
\newblock Side-channel attacks on optane persistent memory.
\newblock In {\em 32nd USENIX Security Symposium (USENIX Security 23)}. USENIX
  Association, August 2023.

\bibitem{dataset:keystroke}
Chen~Change Loy.
\newblock Keystroke100 dataset.
\newblock
  \url{http://personal.ie.cuhk.edu.hk/~ccloy/downloads_keystroke100.html},
  March 2014.

\bibitem{firefox:indexeddb}
{MDN contributors}.
\newblock Using {IndexedDB}.
\newblock
  \url{https://developer.mozilla.org/en-US/docs/Web/API/IndexedDB_API/Using_IndexedDB},
  2023.

\bibitem{db:rocksdb}
{Meta Platforms, Inc.}
\newblock {RocksDB}.
\newblock \url{https://rocksdb.org}, 2022.

\bibitem{gpu:renders}
Hoda Naghibijouybari, Ajaya Neupane, Zhiyun Qian, and Nael Abu-Ghazaleh.
\newblock Rendered insecure: {GPU} side channel attacks are practical.
\newblock In {\em Proceedings of the 2018 ACM SIGSAC Conference on Computer and
  Communications Security}, CCS '18, page 2139–2153, New York, NY, USA, 2018.
  Association for Computing Machinery.

\bibitem{gameserver:Tsung}
Nicolas Niclausse.
\newblock Tsung.
\newblock \url{http://tsung.erlang-projects.org}, 2017.

\bibitem{FS:rethink-sync:OSDI-2006}
Edmund~B. Nightingale, Kaushik Veeraraghavan, Peter~M. Chen, and Jason Flinn.
\newblock Rethink the sync.
\newblock In {\em Proceedings of the 7th Symposium on Operating Systems Design
  and Implementation}, OSDI '06, page 1–14, USA, 2006. USENIX Association.

\bibitem{FS:exF2FS:FAST-2022}
Joontaek Oh, Sion Ji, Yongjin Kim, and Youjip Won.
\newblock {exF2FS}: Transaction support in log-structured filesystem.
\newblock In {\em 20th USENIX Conference on File and Storage Technologies (FAST
  22)}, pages 345--362, Santa Clara, CA, February 2022. USENIX Association.

\bibitem{Chromium:indexeddb}
Jeremy Orlow.
\newblock {IndexedDB} design doc.
\newblock \url{https://www.chromium.org/developers/design-documents/indexeddb},
  2023.

\bibitem{security:Prime+Probe:CT-RSA-2006}
Dag~Arne Osvik, Adi Shamir, and Eran Tromer.
\newblock Cache attacks and countermeasures: The case of {AES}.
\newblock In {\em Proceedings of the 2006 The Cryptographers' Track at the RSA
  Conference on Topics in Cryptology}, CT-RSA'06, page 1–20, Berlin,
  Heidelberg, 2006. Springer-Verlag.

\bibitem{journal:iJournaling:ATC-2017}
Daejun Park and Dongkun Shin.
\newblock {iJournaling}: {Fine-Grained} journaling for improving the latency of
  fsync system call.
\newblock In {\em 2017 USENIX Annual Technical Conference (USENIX ATC 17)},
  pages 787--798, Santa Clara, CA, July 2017. USENIX Association.

\bibitem{FS:msync:Eurosys-2013}
Stan Park, Terence Kelly, and Kai Shen.
\newblock Failure-atomic {Msync()}: A simple and efficient mechanism for
  preserving the integrity of durable data.
\newblock In {\em Proceedings of the 8th ACM European Conference on Computer
  Systems}, EuroSys '13, page 225–238, New York, NY, USA, 2013. Association
  for Computing Machinery.

\bibitem{security:Prime+Scope:CCS-2021}
Antoon Purnal, Furkan Turan, and Ingrid Verbauwhede.
\newblock {Prime+Scope}: Overcoming the observer effect for high-precision
  cache contention attacks.
\newblock In {\em Proceedings of the 2021 ACM SIGSAC Conference on Computer and
  Communications Security}, CCS '21, page 2906–2920, New York, NY, USA, 2021.
  Association for Computing Machinery.

\bibitem{JIT:DeJITLeak}
Qi~Qin, JulianAndres JiYang, Fu~Song, Taolue Chen, and Xinyu Xing.
\newblock {DeJITLeak}: Eliminating {JIT}-induced timing side-channel leaks.
\newblock In {\em Proceedings of the 30th ACM Joint European Software
  Engineering Conference and Symposium on the Foundations of Software
  Engineering}, ESEC/FSE 2022, page 872–884, New York, NY, USA, 2022.
  Association for Computing Machinery.

\bibitem{FS:fsync-failure:ATC-2020}
Anthony Rebello, Yuvraj Patel, Ramnatthan Alagappan, Andrea~C. Arpaci-Dusseau,
  and Remzi~H. Arpaci-Dusseau.
\newblock Can applications recover from fsync failures?
\newblock In {\em Proceedings of the 2020 USENIX Conference on Usenix Annual
  Technical Conference}, USENIX ATC'20, USA, 2020. USENIX Association.

\bibitem{redhat:writeback}
{Red Hat}.
\newblock 5.3.9.5. /proc/sys/vm/.
\newblock
  \url{https://access.redhat.com/documentation/en-us/red_hat_enterprise_linux/4/html/reference_guide/s3-proc-sys-vm},
  2023.

\bibitem{FS:Btrfs}
Ohad Rodeh, Josef Bacik, and Chris Mason.
\newblock {BTRFS}: The linux b-tree filesystem.
\newblock {\em ACM Trans. Storage}, 9(3), aug 2013.

\bibitem{FS:fast-commit-Ext4}
Marta Rybczy\'nska.
\newblock Fast commits for {Ext4}.
\newblock \url{https://lwn.net/Articles/842385/}, January 2021.

\bibitem{10.1145/3366423.3380124}
Iskander Sanchez-Rola, Davide Balzarotti, Christopher Kruegel, Giovanni Vigna,
  and Igor Santos.
\newblock Dirty clicks: A study of the usability and security implications of
  click-related behaviors on the web.
\newblock In {\em Proceedings of The Web Conference 2020}, WWW '20, page
  395–406, New York, NY, USA, 2020. Association for Computing Machinery.

\bibitem{web:Clock}
Iskander Sanchez-Rola, Igor Santos, and Davide Balzarotti.
\newblock Clock around the clock: Time-based device fingerprinting.
\newblock In {\em Proceedings of the 2018 ACM SIGSAC Conference on Computer and
  Communications Security}, CCS '18, page 1502–1514, New York, NY, USA, 2018.
  Association for Computing Machinery.

\bibitem{security:SQLite-query:Security-2021}
Aria Shahverdi, Mahammad Shirinov, and Dana Dachman-Soled.
\newblock Database reconstruction from noisy volumes: A cache side-channel
  attack on {SQLite}.
\newblock In {\em 30th USENIX Security Symposium (USENIX Security 21)}, San
  Diego, CA, August 2021. USENIX Association.

\bibitem{channel:Shannon}
C.~E. Shannon.
\newblock A mathematical theory of communication.
\newblock {\em The Bell System Technical Journal}, 27(3):379--423, 1948.

\bibitem{security:page-fault:ASIACCS-2016}
Shweta Shinde, Zheng~Leong Chua, Viswesh Narayanan, and Prateek Saxena.
\newblock Preventing page faults from telling your secrets.
\newblock In {\em Proceedings of the 11th ACM on Asia Conference on Computer
  and Communications Security}, ASIA CCS '16, pages 317--328, New York, NY,
  USA, 2016. Association for Computing Machinery.

\bibitem{security:browser:USENIX-2021}
Anatoly Shusterman, Ayush Agarwal, Sioli O'Connell, Daniel Genkin, Yossi Oren,
  and Yuval Yarom.
\newblock {Prime+Probe} 1, {JavaScript} 0: Overcoming browser-based
  side-channel defenses.
\newblock In {\em 30th USENIX Security Symposium (USENIX Security 21)}, pages
  2863 -- 2880, San Diego, CA, August 2021. USENIX Association.

\bibitem{FS:journaling-cores:FAST-2018}
Yongseok Son, Sunggon Kim, Heon~Young Yeom, and Hyuck Han.
\newblock High-performance transaction processing in journaling file systems.
\newblock In {\em Proceedings of the 16th USENIX Conference on File and Storage
  Technologies}, FAST'18, page 227–240, USA, 2018. USENIX Association.

\bibitem{keystroke:ssh}
Dawn~Xiaodong Song, David Wagner, and Xuqing Tian.
\newblock Timing analysis of keystrokes and timing attacks on {SSH}.
\newblock In {\em Proceedings of the 10th Conference on USENIX Security
  Symposium - Volume 10}, SSYM'01, USA, 2001. USENIX Association.

\bibitem{auditing:Osquery}
{The Linux Foundation}.
\newblock “osquery.
\newblock \url{https://www.osquery.io}, 2023.

\bibitem{security:no-page-fault:Security-2017}
Jo~Van~Bulck, Nico Weichbrodt, R\"{u}diger Kapitza, Frank Piessens, and Raoul
  Strackx.
\newblock Telling your secrets without page faults: Stealthy page table-based
  attacks on enclaved execution.
\newblock In {\em Proceedings of the 26th USENIX Conference on Security
  Symposium}, SEC'17, page 1041–1056, USA, 2017. USENIX Association.

\bibitem{security:cloud:Security-2015}
Venkatanathan Varadarajan, Yinqian Zhang, Thomas Ristenpart, and Michael Swift.
\newblock A placement vulnerability study in multi-tenant public clouds.
\newblock In {\em Proceedings of the 24th USENIX Conference on Security
  Symposium}, SEC'15, page 913–928, USA, 2015. USENIX Association.

\bibitem{web:defense}
Tao Wang, Xiang Cai, Rishab Nithyanand, Rob Johnson, and Ian Goldberg.
\newblock Effective attacks and provable defenses for website fingerprinting.
\newblock In {\em Proceedings of the 23rd USENIX Conference on Security
  Symposium}, SEC'14, page 143–157, USA, 2014. USENIX Association.

\bibitem{security:page-fault:CCS-2017}
Wenhao Wang, Guoxing Chen, Xiaorui Pan, Yinqian Zhang, XiaoFeng Wang, Vincent
  Bindschaedler, Haixu Tang, and Carl~A. Gunter.
\newblock Leaky cauldron on the dark land: Understanding memory side-channel
  hazards in {SGX}.
\newblock In {\em Proceedings of the 2017 ACM SIGSAC Conference on Computer and
  Communications Security}, CCS '17, page 2421–2434, New York, NY, USA, 2017.
  Association for Computing Machinery.

\bibitem{security:NVLeak:USENIX-2023}
Zixuan Wang, Mohammadkazem Taram, Daniel Moghimi, Steven Swanson, Dean Tullsen,
  and Jishen Zhao.
\newblock {NVLeak}: Off-chip side-channel attacks via non-volatile memory
  systems.
\newblock In {\em 32nd USENIX Security Symposium (USENIX Security 23)}. USENIX
  Association, August 2023.

\bibitem{power:iknowwhatyousee}
Lingxiao Wei, Bo~Luo, Yu~Li, Yannan Liu, and Qiang Xu.
\newblock I know what you see: Power side-channel attack on convolutional
  neural network accelerators.
\newblock In {\em Proceedings of the 34th Annual Computer Security Applications
  Conference}, ACSAC '18, page 393–406, New York, NY, USA, 2018. Association
  for Computing Machinery.

\bibitem{FS:BarrierFS:FAST-2018}
Youjip Won, Jaemin Jung, Gyeongyeol Choi, Joontaek Oh, Seongbae Son, Jooyoung
  Hwang, and Sangyeun Cho.
\newblock Barrier-enabled {IO} stack for flash storage.
\newblock In {\em Proceedings of the 16th USENIX Conference on File and Storage
  Technologies}, FAST'18, page 211–226, USA, 2018. USENIX Association.

\bibitem{security:browser:Security-2022}
Shujiang Wu, Jianjia Yu, Min Yang, and Yinzhi Cao.
\newblock Rendering contention channel made practical in web browsers.
\newblock In {\em 31st USENIX Security Symposium (USENIX Security 22)}, pages
  3183--3199, Boston, MA, August 2022. USENIX Association.

\bibitem{kv:LSM-Trie:ATC-2015}
Xingbo Wu, Yuehai Xu, Zili Shao, and Song Jiang.
\newblock {LSM-Trie}: {An LSM}-tree-based ultra-large key-value store for small
  data.
\newblock In {\em Proceedings of the 2015 USENIX Conference on Usenix Annual
  Technical Conference}, USENIX ATC '15, page 71–82, USA, 2015. USENIX
  Association.

\bibitem{security:page-fault:SP-2015}
Yuanzhong Xu, Weidong Cui, and Marcus Peinado.
\newblock Controlled-channel attacks: Deterministic side channels for untrusted
  operating systems.
\newblock In {\em 2015 IEEE Symposium on Security and Privacy}, pages 640--656,
  2015.

\bibitem{security:directory:SP-2019}
Mengjia Yan, Read Sprabery, Bhargava Gopireddy, Christopher Fletcher, Roy
  Campbell, and Josep Torrellas.
\newblock Attack directories, not caches: Side channel attacks in a
  non-inclusive world.
\newblock In {\em 2019 IEEE Symposium on Security and Privacy (SP)}, pages
  888--904, 2019.

\bibitem{security:cache-coherence:HPCA-2018}
Fan Yao, Milos Doroslovacki, and Guru Venkataramani.
\newblock Are coherence protocol states vulnerable to information leakage?
\newblock In {\em 2018 IEEE International Symposium on High Performance
  Computer Architecture (HPCA)}, pages 168--179, Vienna, Austria, February
  2018. IEEE Press.

\bibitem{security:Flush+Reload:Security-2014}
Yuval Yarom and Katrina Falkner.
\newblock {FLUSH+RELOAD}: A high resolution, low noise, {L3} cache side-channel
  attack.
\newblock In {\em 23rd USENIX Security Symposium (USENIX Security 14)}, pages
  719--732, San Diego, CA, August 2014. USENIX Association.

\bibitem{security:CacheBleed:CHS-2016}
Yuval Yarom, Daniel Genkin, and Nadia Heninger.
\newblock {CacheBleed}: A timing attack on {OpenSSL} constant time {RSA}.
\newblock In Benedikt Gierlichs and Axel~Y. Poschmann, editors, {\em
  Cryptographic Hardware and Embedded Systems -- CHES 2016}, pages 346--367,
  Berlin, Heidelberg, 2016. Springer Berlin Heidelberg.

\bibitem{keystroke:multi-user}
Kehuan Zhang and XiaoFeng Wang.
\newblock Peeping {Tom} in the neighborhood: Keystroke eavesdropping on
  multi-user systems.
\newblock In {\em Proceedings of the 18th Conference on USENIX Security
  Symposium}, page 17–32, USA, 2009. USENIX Association.

\bibitem{pagewake:Binoculars}
Zirui~Neil Zhao, Adam Morrison, Christopher~W. Fletcher, and Josep Torrellas.
\newblock Binoculars: Contention-based side-channel attacks exploiting the page
  walker.
\newblock In {\em 31st USENIX Security Symposium (USENIX Security 22)}, pages
  699--716, Boston, MA, August 2022. USENIX Association.

\bibitem{FS:MAdFS:FAST-2023}
Shawn Zhong, Chenhao Ye, Guanzhou Hu, Suyan Qu, Andrea Arpaci-Dusseau, Remzi
  Arpaci-Dusseau, and Michael Swift.
\newblock {MadFS}: Per-file virtualization for userspace persistent memory
  filesystems.
\newblock In {\em Proceedings of the 21st USENIX Conference on File and Storage
  Technologies}, FAST'23, page 1–15, USA, February 2023. USENIX Association.

\end{thebibliography}

\appendix
\section*{Appendix}
\renewcommand\thefigure{\thesection.\arabic{figure}}    
\renewcommand\thetable{\thesection.\arabic{table}} 
\setcounter{figure}{0} 
\setcounter{table}{0}

\section{Channel Capacity of \Cocoa Channel}\label{sec:app:capacity}

\begin{figure*}[h]
	\centering
	\begin{subfigure}{0.5\textwidth}
		\includegraphics[width=\textwidth]{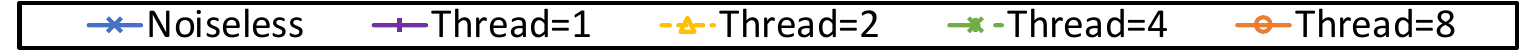}
	\end{subfigure}
	\\
	\begin{subfigure}{0.24\textwidth}
		\includegraphics[width=\textwidth,page=9]{eval-bare.pdf}
		\caption{{\tt fsync}-only within Ext4-Btrfs (w/ and w/o   Fileserver).}
		\label{fig:bare:btrfs-file}
	\end{subfigure}
	\hfill
	\begin{subfigure}{0.24\textwidth}
		\includegraphics[width=\textwidth,page=10]{eval-bare.pdf}
		\caption{{\tt fsync}-only  within Ext4-Btrfs (w/ and w/o  Varmail).}
		\label{fig:bare:btrfs-var}
	\end{subfigure}	
	\hfill
	\begin{subfigure}{0.24\textwidth}
		\includegraphics[width=\textwidth,page=11]{eval-bare.pdf}
		\caption{{\tt write}+{\tt fsync} within Ext4-Btrfs (w/ and w/o   Fileserver).}
		\label{fig:bare:btrfs-w-file}
	\end{subfigure}
	\hfill
	\begin{subfigure}{0.24\textwidth}
		\includegraphics[width=\textwidth,page=12]{eval-bare.pdf}
		\caption{{\tt write}+{\tt fsync} within Ext4-Btrfs (w/ and w/o  Varmail).}
		\label{fig:bare:btrfs-w-var}
	\end{subfigure}
	\hfill
	\begin{subfigure}{0.24\textwidth}
		\includegraphics[width=\textwidth,page=3]{eval-bare.pdf}
		\caption{{\tt write}+{\tt fsync} within Ext4 (w/ and w/o   Fileserver).}
		\label{fig:bare:ext4-w-file}
	\end{subfigure}
	\hfill
	\begin{subfigure}{0.24\textwidth}
		\includegraphics[width=\textwidth,page=4]{eval-bare.pdf}
		\caption{{\tt write}+{\tt fsync} within Ext4  (w/ and w/o  Varmail).}
		\label{fig:bare:ext4-w-var}
	\end{subfigure}
	\hfill
	\begin{subfigure}{0.24\textwidth}
		\includegraphics[width=\textwidth,page=7]{eval-bare.pdf}
		\caption{{\tt write}+{\tt fsync} within Ext4-XFS (w/ and w/o  Fileserver).}
		\label{fig:bare:xfs-w-file}
	\end{subfigure}
	\hfill
	\begin{subfigure}{0.24\textwidth}
		\includegraphics[width=\textwidth,page=8]{eval-bare.pdf}
		\caption{{\tt write}+{\tt fsync} within Ext4-XFS (w/ and w/o  Varmail).}
		\label{fig:bare:xfs-w-var}
	\end{subfigure}
	\caption{The Capacity of Cross-file \Cocoa Channel among Ext4, Btrfs, and XFS.}
	\label{fig:app:bare:channel}
\end{figure*}

\begin{figure}[t]
	\begin{subfigure}{0.23\textwidth}
		\includegraphics[width=\textwidth,page=3]{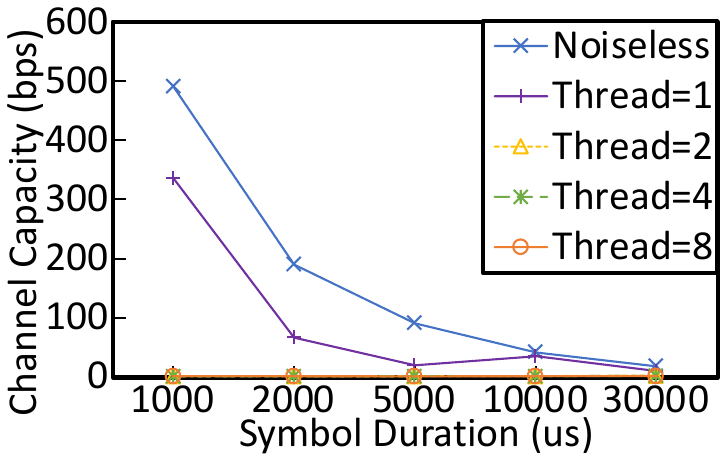}
		\caption{Fileserver Noise.}
		\label{fig:qemu:btrfs-file}
	\end{subfigure}
	\hfill
	\begin{subfigure}{0.23\textwidth}
		\includegraphics[width=\textwidth,page=4]{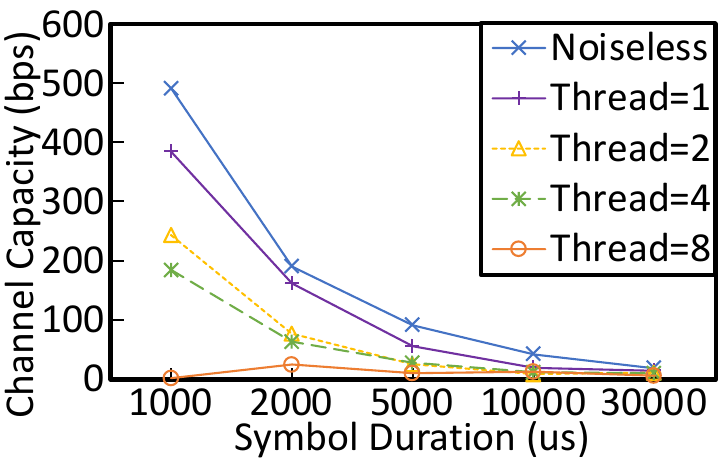}
		\caption{Varmail Noise.}
		\label{fig:qemu:btrfs-var}
	\end{subfigure}
	\caption{The Capacity of Cross-VM \Cocoa Channel between Ext4 and Btrfs.}
	\label{fig:app:qemu:channel}
\end{figure}

\begin{figure}[t]
	\begin{subfigure}{0.23\textwidth}
		\includegraphics[width=\textwidth,page=1]{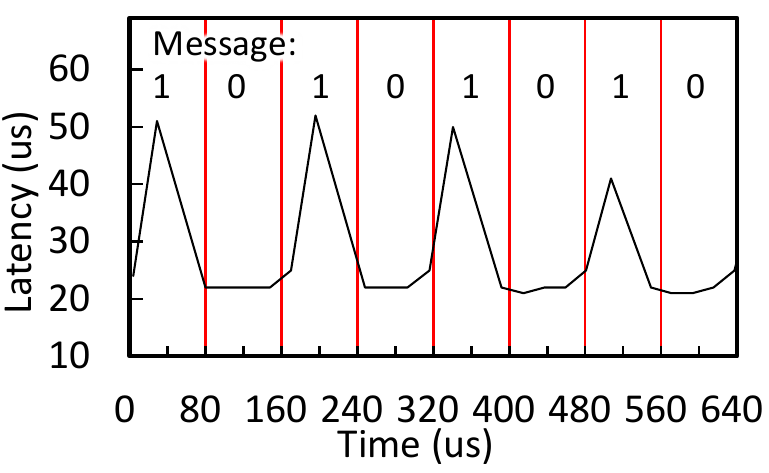}
		\caption{Cross-Container.}
		\label{fig:raw:container}
	\end{subfigure}	
	\hfill
	\begin{subfigure}{0.23\textwidth}
		\includegraphics[width=\textwidth,page=2]{raw-covert.pdf}
		\caption{Cross-VM.}
		\label{fig:raw:vm}
	\end{subfigure}
	\caption{Raw Traces of Cross-Container and Cross-VM Covert Channels with Inter-partition Setting (Ext4-XFS).}
	\label{fig:raw-trace}
\end{figure}

\begin{figure}[t]
	\begin{subfigure}{0.235\textwidth}
		\includegraphics[width=\textwidth,page=1]{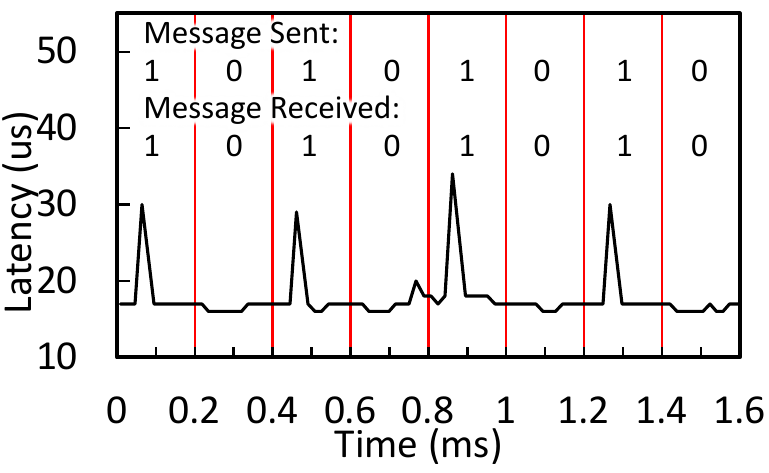}
		\caption{Cross-file (with Fileserver).}
		\label{fig:raw:trace-file-noiseless}
	\end{subfigure}	
	\hfill
	\begin{subfigure}{0.235\textwidth}
		\includegraphics[width=\textwidth,page=2]{raw-noise.pdf}
		\caption{Cross-file (with Varmail).}
		\label{fig:raw:trace-file-noise}
	\end{subfigure}
	\hfil
	\begin{subfigure}{0.235\textwidth}
		\includegraphics[width=\textwidth,page=3]{raw-noise.pdf}
		\caption{Cross-VM (with Fileserver).}
		\label{fig:raw:trace-vm-noiseless}
	\end{subfigure}	
	\hfill
	\begin{subfigure}{0.235\textwidth}
		\includegraphics[width=\textwidth,page=4]{raw-noise.pdf}
		\caption{Cross-VM (with Varmail).}
		\label{fig:raw:trace-vm-noise}
	\end{subfigure}
	\caption{Raw Traces of \Cocoa Covert Channel with Noise caused by Fileserver and Varmail.}
	\label{fig:raw-trace-noise}
\vspace{-1em}

\end{figure}

\begin{figure*}[h]
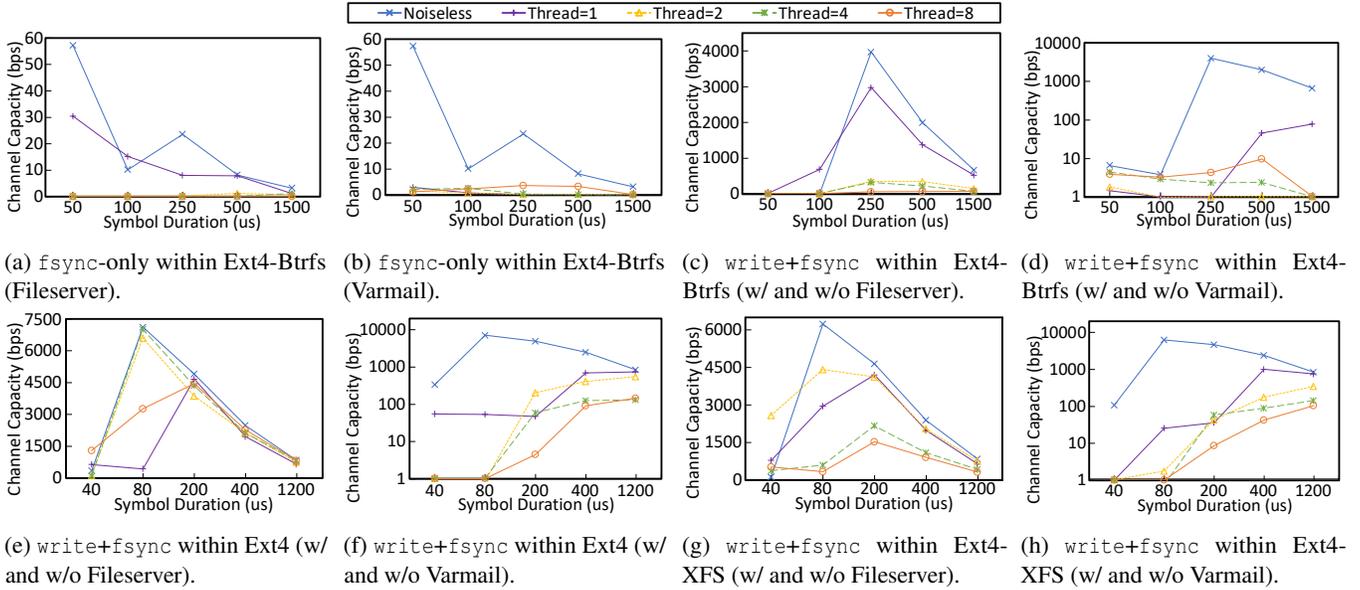

	\centering
	\begin{subfigure}{0.5\textwidth}
		\includegraphics[width=\textwidth]{eval-bar.pdf}
	\end{subfigure}
	\\
	\begin{subfigure}{0.24\textwidth}
		\includegraphics[width=\textwidth,page=9]{eval-docker.pdf}
		\caption{{\tt fsync}-only within Ext4-Btrfs (Fileserver).}
		\label{fig:docker:btrfs-file}
	\end{subfigure}
	\hfill
	\begin{subfigure}{0.24\textwidth}
		\includegraphics[width=\textwidth,page=10]{eval-docker.pdf}
		\caption{{\tt fsync}-only within Ext4-Btrfs (Varmail).}
		\label{fig:docker:btrfs-var}
	\end{subfigure}
	\hfill
	\begin{subfigure}{0.24\textwidth}
		\includegraphics[width=\textwidth,page=11]{eval-docker.pdf}
		\caption{{\tt write}+{\tt fsync} within Ext4-Btrfs (w/ and w/o Fileserver).}
		\label{fig:docker:btrfs-w-file}
	\end{subfigure}
	\hfill
	\begin{subfigure}{0.24\textwidth}
		\includegraphics[width=\textwidth,page=12]{eval-docker.pdf}
		\caption{{\tt write}+{\tt fsync} within Ext4-Btrfs (w/ and w/o Varmail).}
		\label{fig:docker:btrfs-w-var}
	\end{subfigure}	
	\hfill
	\begin{subfigure}{0.24\textwidth}
		\includegraphics[width=\textwidth,page=3]{eval-docker.pdf}
		\caption{{\tt write}+{\tt fsync} within Ext4 (w/ and w/o Fileserver).}
		\label{fig:docker:ext4-w-file}
	\end{subfigure}
	\hfill
	\begin{subfigure}{0.24\textwidth}
		\includegraphics[width=\textwidth,page=4]{eval-docker.pdf}
		\caption{{\tt write}+{\tt fsync} within Ext4 (w/ and w/o Varmail).}
		\label{fig:docker:ext4-w-var}
	\end{subfigure}
	\hfill
	\begin{subfigure}{0.24\textwidth}
		\includegraphics[width=\textwidth,page=7]{eval-docker.pdf}
		\caption{{\tt write}+{\tt fsync} within Ext4-XFS (w/ and w/o Fileserver).}
		\label{fig:docker:xfs-w-file}
	\end{subfigure}
	\hfill
	\begin{subfigure}{0.24\textwidth}
		\includegraphics[width=\textwidth,page=8]{eval-docker.pdf}
		\caption{{\tt write}+{\tt fsync} within Ext4-XFS (w/ and w/o Varmail).}
		\label{fig:docker:xfs-w-var}
	\end{subfigure}
	\caption{The Capacity of Cross-container \Cocoa Channel among Ext4, Btrfs, and XFS.}
	\label{fig:app:docker:channel}
\vspace{-1em}
\end{figure*}

In addition to the \Cocoa channel with {\tt fsync}-only in Section~\ref{sec:channel:eval}, 
the \Cocoa channel can also be established by the sender writing and invoking {\tt fsync} for her/his file, denoted as {\tt write}+{\tt fsync}. 
Detailed results about the capacity of \Cocoa   channel 
are shown in~\autoref{fig:app:bare:channel} and~\autoref{fig:app:docker:channel} 
for cross-file and cross-container, respectively.
\autoref{fig:app:qemu:channel} also depicts cross-VM \Cocoa   channel 
between Ext4 and Btrfs.
\autoref{fig:raw-trace} shows
the raw traces of cross-container and -VM channels,  respectively,
based on Ext4-XFS for transmitting a bit stream of `10101010'.
\autoref{fig:raw-trace-noise} provides raw traces of \Cocoa cross-file and cross-VM covert channel 
with noise cause by Fileserver and Varmail.

\begin{table}[t]
    \centering
	\caption{{\tt fsync} Latency with and without Contention between Ext4 and XFS in Different Disks.}
    \label{tab:latency-ext4}
    \resizebox{\columnwidth}{!}{%
    \begin{tabular}{crrrcrrr}
    \hline
    \begin{tabular}[c]{@{}c@{}}Operation for\\ Measurement\\ (Ext4)\end{tabular} & \begin{tabular}[c]{@{}c@{}}Standalone\\ Latency ($ns$)\end{tabular} & \begin{tabular}[c]{@{}c@{}}Standard\\  Deviation\\  (Standalone)\end{tabular} & \begin{tabular}[c]{@{}c@{}}Standard\\  Error\\  (Standalone)\end{tabular} & \begin{tabular}[c]{@{}c@{}}Operation for\\ Competitor\\ (XFS)\end{tabular} & \begin{tabular}[c]{@{}c@{}}Contention\\ Latency ($ns$)\end{tabular} & \begin{tabular}[c]{@{}c@{}} Standard\\ Deviation\\ (Contention)\end{tabular} & \begin{tabular}[c]{@{}c@{}} Standard\\ Error\\ (Contention)\end{tabular} \\ \hline
    \multirow{3}{*}{{\tt ftruncate}+{\tt fsync}} & \multirow{3}{*}{126149.89} & \multirow{3}{*}{5196.48} & \multirow{3}{*}{519.65} & {\tt ftruncate}+{\tt fsync} & 137153.42 & 7582.32 & 758.23 \\ %
     &  &  &  & {\tt write}+{\tt fsync} & 137590.93 & 8724.20 & 872.42 \\ %
     &  &  &  & {\tt fsync}-only & 143103.89 & 6178.59 & 617.86 \\ %
    \multirow{3}{*}{{\tt write}+{\tt fsync}} & \multirow{3}{*}{54009.40} & \multirow{3}{*}{757.91} & \multirow{3}{*}{75.79} & {\tt ftruncate}+{\tt fsync} & 59385.93 & 5190.36 & 519.04 \\ %
     &  &  &  & {\tt write}+{\tt fsync} & 59234.47 & 1907.39 & 190.74 \\ %
     &  &  &  & {\tt fsync}-only & 61648.70 & 4088.88 & 408.89 \\ %
    \multirow{3}{*}{{\tt fsync}-only} & \multirow{3}{*}{21045.51} & \multirow{3}{*}{316.97} & \multirow{3}{*}{31.70} & {\tt ftruncate}+{\tt fsync} & 22456.78 & 2770.05 & 277.00 \\ %
     &  &  &  & {\tt write}+{\tt fsync} & 24262.37 & 4272.32 & 427.23 \\ %
     &  &  &  & {\tt fsync}-only & 22253.03 & 1611.29 & 161.13 \\ \hline
    \end{tabular}%
    }
\end{table}

\begin{figure}[t]
	\begin{subfigure}{0.23\textwidth}
		\includegraphics[width=\textwidth,page=1]{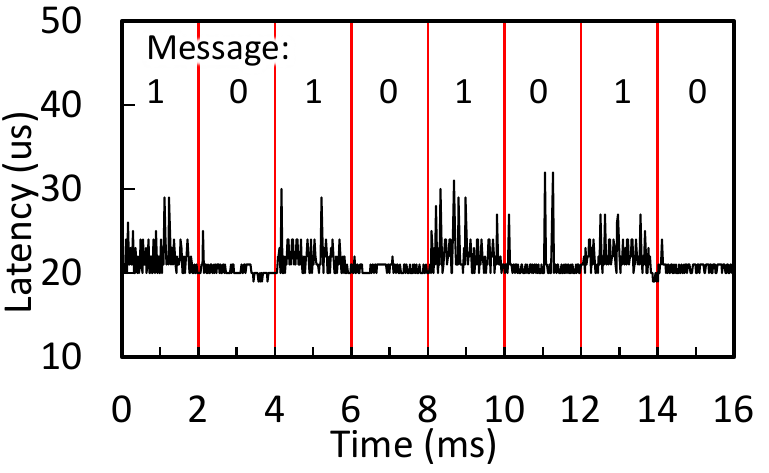}    
		\caption{Cross-file.}
		\label{fig:raw:disk:file}
	\end{subfigure}	
	\hfill
	\begin{subfigure}{0.23\textwidth}
		\includegraphics[width=\textwidth,page=2]{raw-cross.pdf}
		\caption{Cross-VM.}
		\label{fig:raw:disk:vm}
	\end{subfigure}
	\caption{Raw Traces of Cross-disk \Cocoa Covert Channels between Ext4 and XFS.}
	\label{fig:raw-trace-disks}
    \vspace{-1em}
\end{figure}

\section{Additional Discussions}\label{sec:app:discussion}

\subsection{Cross-disk \Cocoa Channel}\label{sec:cross-disk}

We  examine if we can build a reliable cross-disk \Cocoa channel.
We utilize two SSDs (SAMSUNG PM883 SATA SSD) 
and 
build Ext4 and XFS file systems on them, respectively.
Then, we measure {\tt fsync} latencies %
following the methodology and setup %
mentioned in Section~\ref{sec:viability}.
The results are summarized in~\autoref{tab:latency-ext4}.
Interestingly,
although the {\tt fsync} latency with contention only increases by
\fpeval{round(( 24262.37 / 21045.51 - 1)*100, 1)}\% 
compared to that without contention,
the %
standard deviation of latencies
is much higher than that of undisturbed {\tt fsync} latencies.
Take {\tt fsync}-only for instance.
As shown in~\autoref{tab:latency-ext4}, the standard deviations with and without contention
 differ by \fpeval{round(( 2770.05 / 316.97), 1)}$\times$,
\fpeval{round(( 4272.32 / 316.97), 1)}$\times$,
and \fpeval{round(( 1611.29 / 316.97), 1)}$\times$, respectively, under three settings.
This observation indicates the practicability of cross-disk attacks through \Cocoa channel.
We note that the standard deviation of {\tt fsync} latencies with one single disk
drive is not significant. 
The  reason why it substantially varies between SSDs
is on the contention of I/O dispatcher and software queues  
maintained between file system and storage 
device (see Section~\ref{sec:fsync:bio-contention}). 
Given a number of {\tt fsync}s targeting different disks,
the average {\tt fsync} latencies for I/O completion are unlikely to change a lot
because the total time for all {\tt fsync}s is mainly spent 
on storage device, like committing to on-disk journal and competing for using device resources.
I/O dispatcher and software queues, however, determine
when an {\tt fsync} would be delivered to its device. An earlier
dispatched {\tt fsync} to one disk hence hinders another concurrent {\tt fsync}
from going
to the other disk, thereby generating substantial difference (standard deviation) 
of {\tt fsync} latencies.

We %
utilize the standard deviation of {\tt fsync} latencies
as criteria to build  cross-disk \Cocoa channel.
As shown in~\autoref{fig:raw:disk:file}, we set the sender with XFS
that synchronizes its file after writing 1KB file data %
while the receiver stays in the other SSD with Ext4 mounted,
only calling {\tt fsync}s periodically. This
cross-disk channel %
achieves 500bps bandwidth at 0.46\% bit error rate 
in an environment without noise.
The transmission rate is lower than the \Cocoa channel conducted in one disk, 
because the sender now needs to call {\tt fsync}s for enough times so that
the receiver   
can distinguish `0' and `1'
by analyzing standard deviations over multiple sampled latencies.

Next, we try to make cross-VM transmission with cross-disk \Cocoa channel.
We place two disk images in two SSDs with XFS and Ext4 mounted, respectively.
Both sender and receiver write their files before synchronizing files.
Whereas, our ample test cases show that
it is difficult to establish a reliable cross-disk covert channel between VMs.
In the most cases, the error rates 
of sending a frame of 8,000 bits have been over 30\%.
The contention across disks is weaker than those within one disk.
The I/O virtualization done by the hypervisor further reduces such contention.
In fact, 
measured {\tt fsync} latencies in a guest OS are not stable even without any interference from
the other guest.

\begin{table}[t]
    \centering
	\caption{Bandwidth and Raw Bit Error Rate of Cross-file \Cocoa channel on Other Platforms without Noise.}
    \label{tab:res:platforms}
    \resizebox{\columnwidth}{!}{%
    \begin{tabular}{ccccc}
    \hline
    \multirow{2}{*}{Platform} & \multicolumn{2}{c}{\begin{tabular}[c]{@{}c@{}}Sender: {\tt fsync}-only\\ Receiver: {\tt fsync}-only\end{tabular}} & \multicolumn{2}{c}{\begin{tabular}[c]{@{}c@{}}Sender: {\tt write} + {\tt fsync}\\ Receiver: {\tt fsync}-only\end{tabular}} \\ \cline{2-5} 
     & \multicolumn{1}{c}{Bandwidth (bps)} & Bit Error Rate & \multicolumn{1}{c}{Bandwidth (bps)} & Bit Error Rate \\ \hline
    2nd Server & \multicolumn{1}{c}{5,000} & 0.39\% & \multicolumn{1}{c}{333.3} & 6.65\% \\ %
   Workstation & \multicolumn{1}{c}{833.3} & 1.42\% & \multicolumn{1}{c}{500} & 0.57\% \\ \hline
    \end{tabular}%
    }
\vspace{-1em}
\end{table}

However, 
cross-disk \Cocoa covert channel between VMs is occasionally %
 possible. In the best case we have obtained over time, %
\Cocoa could transmit at about 500bps %
 with 15.8\% error rate and no background noise.
 As shown in~\autoref{fig:raw-trace-disks},
a comparison between raw traces for sending `10101010' %
also addresses the volatility of cross-disk \Cocoa channel for cross-VM scenarios.
We leave how to build a stable, robust, and efficient cross-disk \Cocoa channel as one of our future works.

\vspace{-1em}
\subsection{The Impact of Platforms and NVMe SSD}

We test the capability of \Cocoa on multiple platforms. Firstly,
we validate \Cocoa channel in a server with 
Intel Xeon Gold 5218 CPU and 960G Micron 5300 PRO SATA SSD.
Secondly, we  deploy it on NVMe SSDs, say,
1TB SAMSUNG 970 PRO NVMe SSD, and 512GB SK Hynix PC601 NVMe SSD, 
in a workstation with Intel Core i9-9900K CPU.
The OS is  Ubuntu 22.04 for both machines.
\autoref{tab:res:platforms} reveals cross-file \Cocoa channel built on Ext4 in these two platforms.  
In spite of multiple hardware queues in an NVMe SSD, the
contention of calling concurrent {\tt fsync}s still exists. 
The CPUs of these platform are not as powerful as the one of our main platform 
(Intel Xeon Gold 6342).
\Cocoa channels' bandwidth in these platforms is lower than 
 in previous platform mentioned in Section~\ref{sec:setup}, which
 indicates {\tt fsyncs} latency is not only related to storage devices, 
but also influenced by CPUs.
This observation aligns with prior %
 studies~\cite{FS:journaling-cores:FAST-2018,journal:Z-journal:ATC-2021} intended to
exploit multi-core CPU %
to boost performance for file system and newer SSDs. %
In addition, cross-disk \Cocoa channel between files 
achieves 125bps bandwidth with 13.1\% error rate in NVMe SSDs.

\begin{figure}[t]
	\centering
	\includegraphics[width=0.6\columnwidth,page=3]{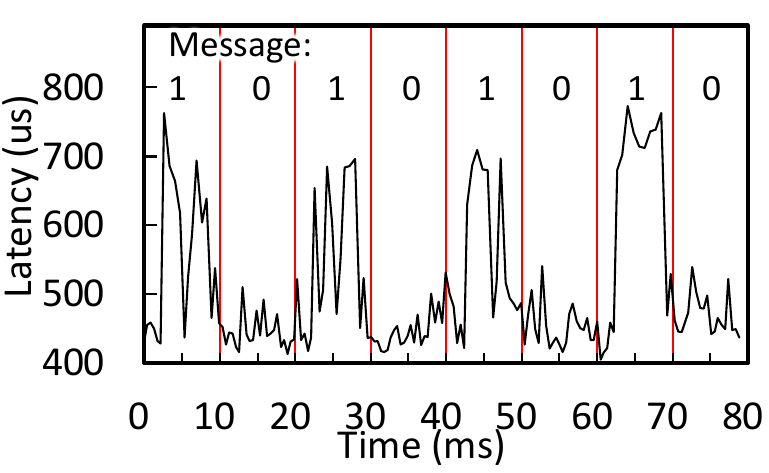}
	\vspace{-1em}
	\caption{Raw Traces of Pass-through Disk in VMs.}
	\label{fig:raw:passthrough}
	\vspace{-1.5em}
\end{figure}

\vspace{-1em}
\subsection{Pass-through Disk in VMs for \Cocoa}

KVM has a {\em Passthrough} feature 
that adds an additional storage device to a guest VM 
in order to provide increased storage space
or separate system data from user data. %
Additionally,
 adding disk partitions to guest VMs with Passthrough is considered
to ensure higher security than adding a whole disk. 
We build cross-VM \Cocoa    channel that is %
established on two pass-through partitions of one disk.
This channel differs from the one shown in Section~\ref{sec:covert:model}
in that the latter has been 
made on two image files, rather than two partitions directly.
We have tested with different compositions of file systems on two partitions. Without loss of generality for illustration, we choose a scenario in which
the sender's   and receiver's VMs are mounted with Btrfs and Ext4, respectively.
Cross-VM \Cocoa   channel functioning on pass-through partitions 
 can transmit at about 100bps
 with 8.28\% error rate with no background noise.
A raw trace for `10101010' is captured by~\autoref{fig:raw:passthrough}.
Compared to the cross-VM attacks with disk image files, the channel capacity of \Cocoa decreases
on pass-through 
partitions.
The reason %
is that the contention across disk partitions is weaker than
 that of sharing and contending in the same file system with disk image files.

\vspace{-0.5em}
\section{Algorithms in \Cocoa}

\begin{algorithm}[t]
	\caption{Sending frame via \Cocoa }\label{alg:sender}
	\KwIn{data frame array $frame$, 
		data frame length $len$, 
		symbol duration $t_s$, 
		file   $fd$}
	\For{$i\leftarrow 0$ \KwTo $len - 1$}{
		\eIf (// send bit `1'.)
		{$frame[i]$ == $1$}
		{
			$run\_time$ = 0\;
			// execute the following operations for $t_s$.\\
			\While {$run\_time$ < $t_s$}
			{
				ftruncate($fd$) or write($fd$) or do nothing in some cases\;
				fsync($fd$)\;
				update $run\_time$\;
			}
		}(// send bit `0'.)
		{
			sleep($t_s$)\;
		}
	}
\end{algorithm}

Algorithms~\ref{alg:sender} and~\ref{alg:receiver} present main steps %
the
sender 
and receiver follow to transmit a bit with the protocol of \Cocoa.

\begin{algorithm}[h]
	\caption{Receiving frame via \Cocoa.} \label{alg:receiver}
	\KwIn{threshold $\theta$, 
		measure time period $t_s$, 
		file   $fd$}
	\While {$true$}
	{
		$run\_time$ = 0\;
		empty($time\_list$)\;
		// execute the following operations for $t_s$.\\
		\While {$run\_time$ < $t_s$}
		{
			ftruncate($fd$) or write($fd$) or do nothing in some cases\;
			record time $t_{start}$\;
			fsync($fd$)\;
			record time $t_{end}$\;
			$time\_list[i++] = t_{end} - t_{start}$\;
			update $run\_time$\;
		}
		\eIf{$avg(time\_list) > \theta$}
		{
			output(`1')\;
		}
		{
			output(`0')\;
		}
	}
\end{algorithm}

\end{document}